\documentclass{emulateapj}

\usepackage{amsmath}
\usepackage{amsfonts}
\usepackage{amssymb}
\usepackage{bm}
\usepackage{cases}
\usepackage{float}
\usepackage{graphicx}
\usepackage{latexsym}
\usepackage{longtable}
\usepackage{multirow}
\usepackage{txfonts}
\usepackage{tipa}

\shorttitle{Halo Stream}
\shortauthors{HeFan Li et al.}

\begin{document}

\title{The substructures in the local stellar halo from Gaia and LAMOST}

\author{Hefan Li\altaffilmark{1}, Cuihua Du\altaffilmark{2,3}, Shuai Liu\altaffilmark{4}, Thomas Donlon \altaffilmark{3}, Heidi Jo Newberg\altaffilmark{3}}

\affil{$^{1}$School of Physical Sciences, University of Chinese Academy of Sciences, Beijing 100049, P. R. China;\\
$^{2}$College of Astronomy and Space Sciences, University of Chinese Academy of Sciences, Beijing 100049, China; ducuihua@ucas.ac.cn\\
$^{3}$Department of Physics, Applied Physics and Astronomy, Rensselaer Polytechnic Institute, Troy, NY 12180, USA\\
$^{4}$Key Laboratory of Optical Astronomy, National Astronomical Observatories, Chinese Academy of Sciences, Beijing 100012, China\\}

\begin{abstract}
\par Based on the second Gaia data release (Gaia DR2) and spectroscopy from the Large Sky Area Multi-Object Fiber Spectroscopic Telescope (LAMOST) Data, we identified 20,089 halo stars kinematically and chemically. The halo streams in the solar neighborhood could be detected in the space of energy and angular momentum. We reshuffle the velocities of these stars to determine the significance of substructure. Finally, we identify 4 statistically significant substructures that are labeled GL-1 through 4. Among these substructures,  GL-1 is previously known stream (``N2" stream)  and 
the rest 3 substructures are new. These substructures may be the debris of dwarf galaxies accretion event, their dynamical and chemical information can help to understand the history of the Milky Way.

\end{abstract}


\section{Introduction}

\par In the standard hierarchical model of galaxy formation, stellar halos are thought to form via the accumulation of stars from dwarf galaxies that accreted and merged with a larger galaxy. This merging process left behind many stellar streams or moving groups in the Galactic halo \citep{Searle78, Freeman02}. Although the accumulated debris from old accretion events rapidly disperses in real space, stellar halos still maintain visible fossil structure in their phase space distribution \citep{Helmi99a,Bullock05}. Nonetheless, many Galactic halo substructures have been spatially identified using data from large sky surveys such as the Sloan Digital Sky Survey \citep[SDSS;][]{York00}, Hipparcos \citep{Leeuwen07}, the Radial Velocity Experiment \citep[RAVE;][]{Steinmetz06} , the Two Micron All-Sky Survey \citep[2MASS;][]{2MASS} and Wide-Field Infrared Survey Explorer \citep[WISE;][]{WISE} surveys. Some of the substructures identified include: the Sagittarius dwarf tidal stream \citep{Ibata94,Majewski03,Belokurov06}, the Orphan Stream \citep{Grillmair06,Belokurov07,Newberg10}　and the Virgo Stellar Stream \citep{Duffau06,Newberg07,Duffau14}. \citet{Grillmair16} summarize the methods used for discovering streams and present tables that give some basic information for most of the streams and clouds found before 2015, and they also discuss the properties of individual tidal debris structures.
 
\par Numerical simulations predict that most accreted satellites would also spread their tidal debris into the Galactic inner halo \citep{Helmi99a} or even the disk component \citep{Abadi03}, thus we can expect to find debris remnants in the local halo. Some efforts have been made to find the debris relics in the local halo and the observational evidence is slowly growing. For example, \citet{Helmi99b} have detected two streams in the solar neighborhood by studying the angular momentum of stellar orbits. \citet{Smith09} found four discrete overdensities localized in angular momentum in the solar neighborhood and suggested that they may be possible accretion remnants. \citet{Kepley07} assembled a sample of halo stars in the solar neighborhood to find two new halo substructure in velocity and angular momentum space. \citet{Morrison09} have also shown that the angular momentum distribution of a sample of stars does not seem to be smooth, possibly indicative of an accretion origin. \citet{Klement10} summarizes the discoveries of solar neighborhood stellar halo streams and gives a theoretical overview over the search strategies employed. Recently, \citet{Helmi17} study the distribution of local halo stars in ``Integrals of Motion'' space using the TGAS dataset, in combination with data from the RAVE survey and discover several substructures that could potentially be related to merger events. Although no longer spatially coherent, such stellar streams keep their common origin imprinted into their chemical and dynamical properties. This allows us constrain various scenarios of hierarchical buildup of the Galaxy \citep[e.g.,][]{Helmi99a} from the phase-space distribution of local halo stars.
 
\par The chemical evolution of stellar halos \citep{Tissera14}, supports the idea that the outer halo formed primarily through the accretion of smaller stellar systems like dwarf spheroidal (dSph) galaxies \citep{Searle78,Sales07,Diemand08,Springel08,Klypin11}. On the other hand, part of the inner halo may have formed in-situ, either from dissipative collapse of gaseous material onto the central region of the Galaxy \citep{Eggen62, Zolotov10, Font11}, or from a heated disk during merger events \citep{Cooper15}. Evidence for the dual halo (either inner vs. outer, or metal-rich vs. metal-poor) has been found from the kinematics of stars near the Sun by \citet{Chiba00,Carollo07,Carollo10,Deason17,Belokurov18}. It is possible that the inner-halo was also built by accretion.

\par Thanks to modern sky surveys that provide radial velocities and proper motions, we can use the full 6D phase space information for halo stars to search for substructures in the solar neighborhood. This will allow us to search for older substructures that could have lost their spatial coherence but maintain coherence in energy and angular momentum.

\par ESA's Gaia mission has produced the richest star catalogue to date, including high-precision measurements of nearly 1.7 billion stars \citep{Gaia18}, in April 2018. Gaia DR2 included positions, distance indicators and motions of more than one billion stars. To obtain the required full phase-space information of halo stars, we also use radial velocities and metallicities derived from Large Sky Area Multi-Object Fiber Spectroscopic Telescope \citep[LAMOST;][]{Cui12,Deng12,Zhao12}. In July 2017, LAMOST release its the Data 5 (LAMOST DR5), which includes over 9 million spectra. In this paper, we will use the combined data set obtained by cross-matching Gaia to LAMOST to find the potential substructures.

\par The paper is organized as follows. In Section \ref{data}, we introduces the observational data from Gaia and LAMOST, describe the sample selection, and define the coordinate systems used in this study. The detection strategy and analysis of detected substructures are discussed in Section \ref{analysis}. The conclusions are summarized in Section \ref{conclusion}.

\section{DATA}
\label{data}

\subsection{Gaia and LAMOST}

\par The Gaia satellite is a space-based mission launched at the end of 2013 and started science operations the following year. The second Gaia data release (Gaia DR2) includes high-precision measurements of nearly 1.7 billion stars \citep{Gaia18}. As well as positions, the data include astrometry, photometry, radial velocities, and information on astrophysical parameters and variability, for sources brighter than magnitude 21. This data set contains positions, parallaxes, and mean proper motions for about 1.3 billion of the brightest stars. For a subset of stars within a few thousand parsec of the Sun, Gaia has measured the velocity in all three dimensions.

\par The Large Sky Area Multi-Object Fiber Spectroscopic Telescope (LAMOST, also called the Guo Shou Jing Telescope) is a 4 meter quasi-meridian reflective Schmidt telescope with 4000 fibers within a field of view of $5^{\circ}$. The LAMOST spectrograph has a resolution of R $\rm \sim$ 1,800 and wavelength range spanning 3,700 {\AA} to 9,000 {\AA} \citep{Cui12}. LAMOST has completed 5 years of survey operations plus a Pilot Survey, and has internally released over 9 million spectra to the collaboration. The survey reaches a limiting magnitude of $r=17.8$ (where $r$ denotes magnitude in the SDSS $r$-band), but most targets are brighter than $r\sim17$. The LAMOST Stellar Parameter Pipeline \citep{Wu11,Luo15} estimates parameters, including radial velocity, effective temperature, surface gravity and metallicity ([Fe/H]) from LAMOST spectra. The accuracy depends on wavelength calibration, spectral type and SNR of spectra. The accuracies in measuring radial velocity and [Fe/H] at R = 1800 are expected to be 7 km s$^{-1}$ and 0.1 dex, respectively \citep{Zhao12}. In total, there are over 5 million stars in the A, F, G and K type star catalog.

\par The estimated systematic offset of Gaia DR2 parallaxes is $-0.029$ based on quasar \citep{Lindegren18}. Using the catalogue of radial velocity standard stars provided by \citet{Huang18}, we determine the radial velocity zero-points (RVZPs) of LAMOST DR5 is $\Delta$RV = $-4.70$ kms$^{-1}$. The parallax and radial velocity measurements are corrected with their offsets in the following study. Our initial sample was obtained by cross-matching between the Gaia and LAMOST catalogs, based on stellar position.  After imposing the requirements that relative parallax error $\leq 20\%$, the signal-to-noise ration (SNR) $\geq 20$ and radial velocity error $\epsilon_{\mathrm{RV}} \leq$ 10 km s$^{-1}$, the data set contains 2,641,631 stars.

\subsection{Coordinate transformations}  

\par To get full 6D phase-space information, we transform the coordinates measured for the stars into a Cartesian coordinate system. 

\par Parallaxes are one of the few distance measures in astronomy which do not require assumptions about the intrinsic properties of the object. Despite the simple relation between the parallax and distance, inversion of the parallax to obtain distance is only appropriate when there are no measurement errors.  So we use the full Bayesian approach to infer distances and velocities \citep{Luri18}. We use the exponentially decreasing space density prior in distance $d$ \citep{Bailer18}:
\begin{equation}
P(d\ |\ L) \propto d^2 \exp (-d/L)
\end{equation}
and assume that the prior of $v_{\alpha}, v_{\delta}, v_r$ are uniform. Then we can write the posterior as:
\begin{equation}
P(\bm{\theta}\ |\ \bm{x}) \propto \exp [-\frac{1}{2} (\bm{x} - \bm{m(\theta)})^\mathrm{T} C_x^{-1} (\bm{x} - \bm{m(\theta)})]\ P(d\ |\ L)
\end{equation}
where $\bm{\theta} = (d,\ v_{\alpha},\ v_{\delta},\ v_r)^\mathrm{T}$, $\bm{x} = (\varpi,\ \mu_{\alpha^*},\ \mu_{\delta},\ rv)^\mathrm{T}$, $\bm{m} = (1/d,\ v_{\alpha}/kd, \ v_{\delta}/kd,\ v_r)^\mathrm{T}$, $k$ = 4.74 and $C_x^{-1}$ is covariance matrix. We choose the most probable value of $d, v_{\alpha}, v_{\delta}, v_r$ in the posterior as the estimator of the distances and velocities.

\par We calculate the Galactocentric Cartesian $(x, y, z)$ coordinates \citep{Juric08} from the Galactic $(l, b)$ coordinates and distances as follows:
\begin{equation}
\begin{split}
&x = R_{\odot} - d \cos(b) \cos(l)\\
&y = - d \cos(b) \sin(l)\\
&z = d \sin(b) + z_{\odot}\\
\end{split}
\end{equation}
where $R_{\odot}$ = 8.2 kpc is the distance to the Galactic center \citep{Bland16}, $z_{\odot}$ = 25 pc is the solar offset from local disk midplane \citep{Juric08}, $d$ is distance of the star from the Sun, and ($l$, $b$) are the Galactic coordinates. We calculate each star's Galactic space-velocity components, $U$, $V$ and $W$, from its $d, v_{\alpha}, v_{\delta}, v_r$ mentioned above \citep{Johnson87}. We correct for the solar peculiar motion of $(U, V, W) = (10., 11., 7.)$ km s$^{-1}$ \citep{Tian15, Bland16} relative to the local standard of rest (LSR). Here, we assume that the LSR velocity is $V_\mathrm{{LSR}}$ = 232.8 km s$^{-1}$ in the direction of rotation \citep{McMillan17}.

\par We calculate the cylindrical velocities:
\begin{equation}
\begin{split}
&v_R = -U\cos(\phi) - V\sin(\phi)\\
&v_{\phi} = U\sin(\phi) - V\cos(\phi)\\
&v_z = -W\\
\end{split}
\end{equation}
where $\phi = \tan^{-1}(y/x)$, and the spherical velocities for our sample:
\begin{equation}
\begin{split}
&v_r = -U\sin(\theta)\cos(\phi) - V\sin(\theta)\sin(\phi) - W\cos(\theta)\\
&v_{\theta} = -U\cos(\theta)\cos(\phi) - V\cos(\theta)\sin(\phi) + W\sin(\theta)\\
&v_{\phi} = U\sin(\phi) - V\cos(\phi)\\
\end{split}
\end{equation}
where $r = \sqrt{x^2 + y^2 + z^2}, \theta = \cos^{-1}(z/r)$ and $\phi = \tan^{-1}(y/x)$.

\begin{figure*}[t]
	\centering
	\includegraphics[width=1.0\textwidth]{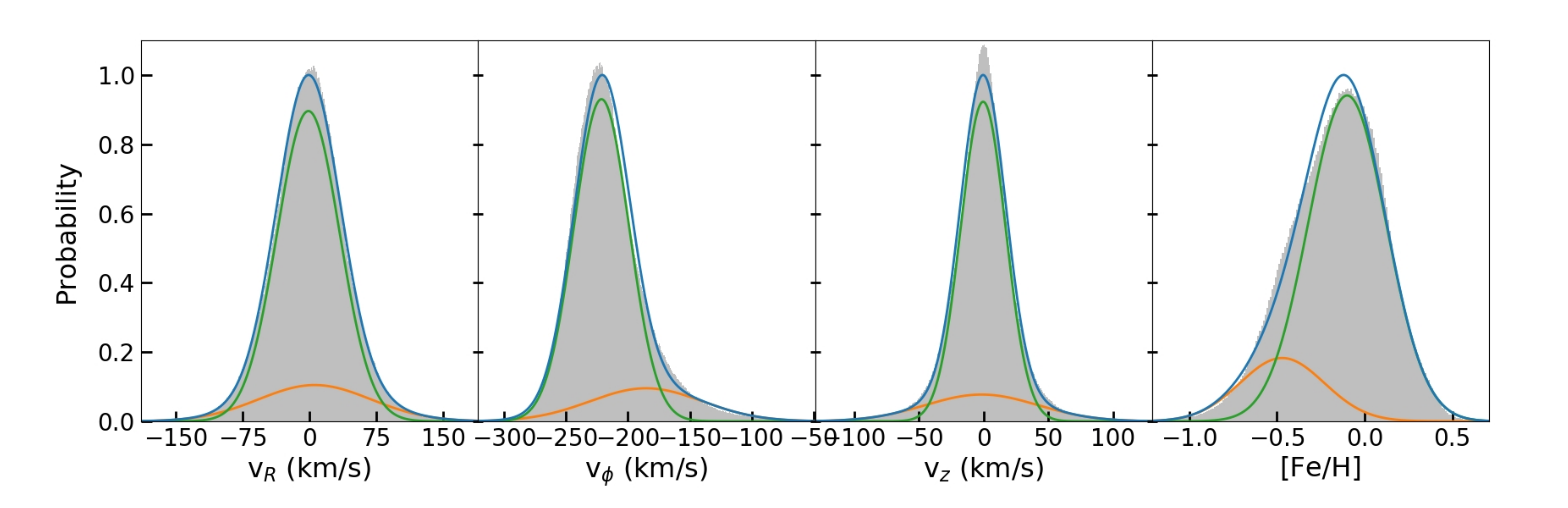}
	\caption{Velocity and chemical distribution of disk-classified star cross-matched between Gaia and LAMOST. The observation is shown as gray histogram. The green and orange line represent the thin disk and the thick disk. These components were identified using a two-component Gaussian model. And the blue line shows the total model.}
	\label{gmm}
\end{figure*}

\subsection{Decomposition into disk and halo}

\par Selecting halo stars based on metallicity is a common method. Disk stars are metal-rich, with a peak in the metallicity distribution at approximately solar metallicity, [Fe/H] = 0. The halo is more metal poor; the inner halo has a typical metallicity of [Fe/H]$=-1.45$ \citep[e.g.,][]{Zuo17} or [Fe/H] $= -1.6$ \citep[e.g.,][]{Allende Prieto06}. However, \citet{Bonaca17} use the first Gaia data, combined with the RAVE and APOGEE spectroscopic surveys, to reveal a metal-rich halo component within $\lesssim 3$ kpc from the Sun. They select halo stars kinematically with a relative velocity of at least 220 km s$^{-1}$ with respect to the local standard of rest and find half of their halo sample is comprised of stars with [Fe/H] $> -1$. The metallicity appears to extend even to super-solar values. So we will lose many metal-rich halo stars if we select halo stars by metallicity.

\par Another commonly used selection criterion for halo stars is $\left| \textbf{V} - \textbf{V}_{\mathrm{LSR}} \right| > \textbf{V}_{\mathrm{cut}}$ in Toomre diagram of Cartesian coordinates. However, the velocity dispersion are different and covariance coefficient are not zero in Cartesian coordinates. If adopting the spherical or cylindrical coordinate system,  the diagonal covariance matrix can be applied.

\par In this work, we use the method provided by \citet{Bensby03} to obtain the probabilities that a given star belongs to a specific population by assuming the Galactic space velocities and metallicity of the stellar populations in the thin disk, the thick disk, and the halo have Gaussian distributions.  Halo stars are too few to be identified obviously as a Gaussian component.  So we remove stars with $v_{\phi} > -40$ km s$^{-1}$ and [Fe/H] $< -1.4$. Then we use the $\textsc{sci-kit learn}$ package in $\textsc{python}$ \citep{Pedregosa12} to fit a two-component Gaussian Mixture Model to the spherical velocities and metallicity of the stars, $v_R$, $v_{\phi}$, $v_z$ and [Fe/H]. In this fitting, each component has its own diagonal covariance matrix. The velocity dispersion are $(\sigma_{R}, \sigma_{\phi}, \sigma_{z}) = (35, 22, 17)$ km s$^{-1}$ for the thin disk and $(\sigma_{R}, \sigma_{\phi}, \sigma_{z}) = (63, 45, 43)$ km s$^{-1}$ for the thick disk. The mean of $v_{R}$ and $v_{z}$ are close to 0 km s$^{-1}$ for both thin and thick disk. The mean of $v_{\phi}$ is $-$221 km s$^{-1}$ for the thin disk and $-$185 km s$^{-1}$ for the thick disk. The mean and dispersion of metallicity are $-0.10,\ 0.22$ for the thin disk and $-0.47,\ 0.24$ for the thick disk.  The fitting result is shown in Figure \ref{gmm} and we get the probability that star belongs to thin disk (D) or thick disk (TD).

\par \citet{Bond10} used SDSS data to analyze velocities of about 100 thousand halo stars. They found the mean rotation of halo stars is $\mu_{\phi} = -(V_{\mathrm{LSR}} + V_{\odot} - 205)$ km s$^{-1}$ = $-$38.8 km s$^{-1}$, where $V_{LSR}$ = 232.8 km s$^{-1}$ and $V_{\odot}$ = 11 km s$^{-1}$. And the velocity ellipsoid $(\sigma_{r}, \sigma_{\theta}, \sigma_{\phi}) = (141, 75, 85)$ km s$^{-1}$ whose principal axes align well with spherical coordinates.  Based on the SDSS and SCUSS, \citep{Zuo17} found the mean metallicity of inner halo is $\mu_{\mathrm{[Fe/H]}} = -1.43$ and the dispersion is $\sigma_{\mathrm{[Fe/H]}} = 0.36$. So we can use the four-dimensional Gaussian distribution, i.e.

\begin{equation}
\begin{split}
&f(v_{r},\, v_{\theta},\, v_{\phi},\, \mathrm{[Fe/H]}) =\\
&k \cdot
\exp \left( -\frac{v_r^2}{2 \sigma_r^2} - \frac{v_{\theta}^2}{2 \sigma_{\theta}^2} - \frac{(v_{\phi}-\mu_{\phi})^2}{2 \sigma_{\phi}^2}
- \frac{(\mathrm{[Fe/H]}-\mu_{\mathrm{[Fe/H]}})^2}{2 \sigma_{\mathrm{[Fe/H]}}^2} \right)\\
\end{split}
\end{equation}

\begin{figure}
	\centering
	\includegraphics[width=1.0\hsize]{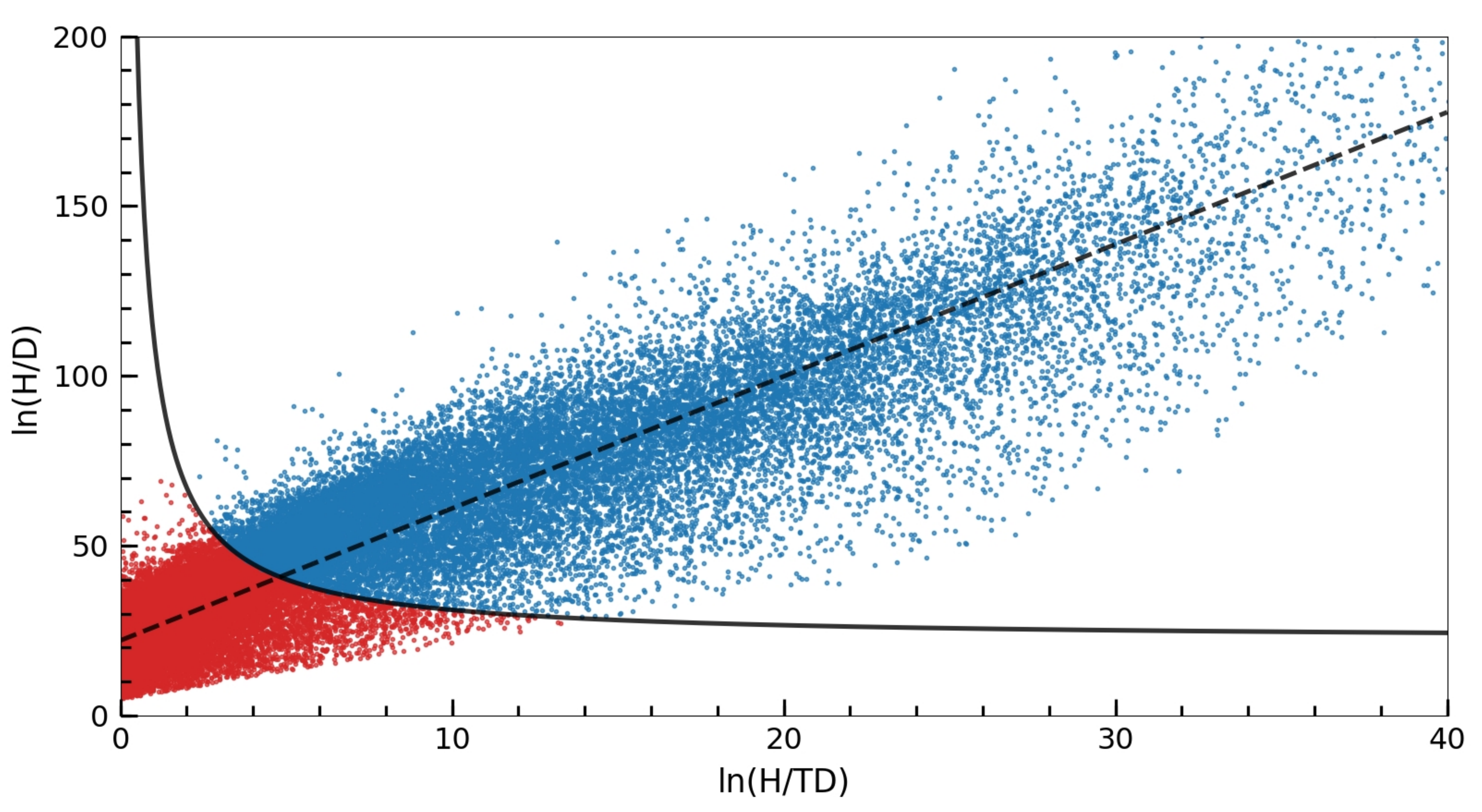}
	\caption{Distribution of $ln(\mathrm{H/D})$ vs $ln(\mathrm{H/TD})$ for the stars with H$>$D and H$>$TD. The black dash line is the linear function for fitting $ln(\mathrm{H/TD})$ and $ln(\mathrm{H/D})$ using least square method. The halo stars (blue dots) are defined as having $ln(\mathrm{H/TD})*(ln(\mathrm{H/D})-b) >90$, where $b=22.10$ is the y-intercept and the dividing line is shown in solid black.}
	\label{HDTD}
\end{figure}

\begin{figure*}[t]
	\centering
	\includegraphics[width=0.8\textwidth]{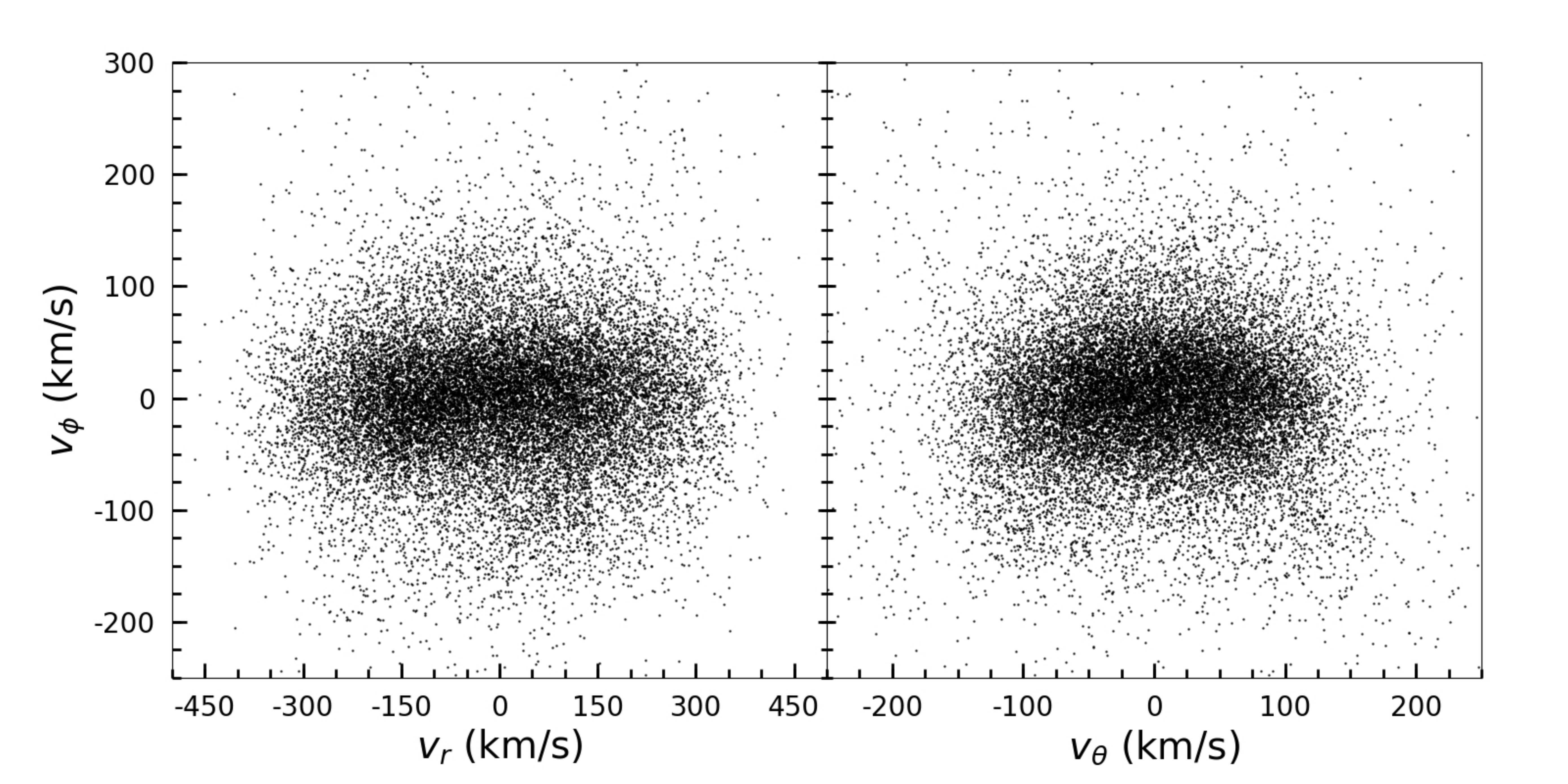}
	\caption{Velocities distribution of our halo sample. These stars are classified as belonging to the halo according to a four-component Gaussian model.}
	\label{sample}
\end{figure*}

\begin{figure}
	\centering
	\includegraphics[width=1.0\hsize]{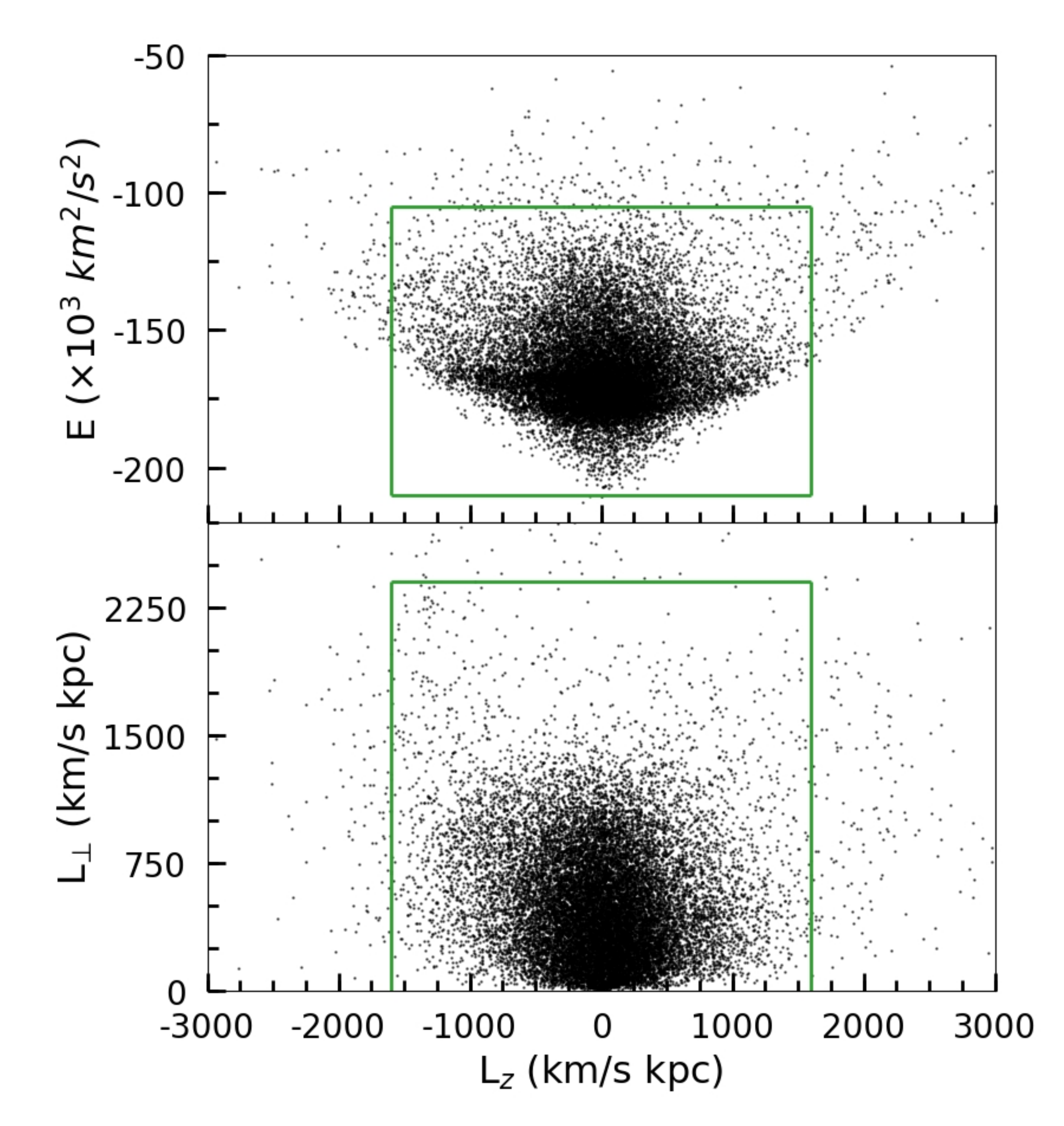}
	\caption{ Distribution of Energy vs $L_z$ (top panel), and $L_{\bot}$ vs $L_z$ (bottom panel) for the halo sample that was selected in velocity and metallicity from the stars that were in common to both Gaia and LAMOST. The green rectangle shows range of Figure \ref{kde}. }
	\label{ELL}
\end{figure}

where $k = \frac{1}{(2\pi)^2\,\sigma_r\, \sigma_{\theta}\, \sigma_{\phi}\, \sigma_{\mathrm{[Fe/H]}}}$, to get the probability that star belongs to halo (which we call H). Now we can derive two relative probabilities for the halo-to-thin-disk (H/D) and halo-to-thick-disk (H/TD) for each star in our sample.  Because of the similarities between the thin disk and the thick disk, when one of $ln(\mathrm{H/TD})$ and $ln(\mathrm{H/D})$ is large, the other is generally large. So after selecting stars with H$>$D and H$>$TD, the least square method is applied to fit line and get the linear relation between $ln(\mathrm{H/TD})$ and $ln(\mathrm{H/D})$:
\begin{equation}
ln(\mathrm{H/D}) = 3.89*ln(\mathrm{H/TD}) + 22.10
\end{equation}
Then we select sample halo stars with $ln(\mathrm{H/TD})*(ln(\mathrm{H/D})-22.10) >90$. The result is shown in Figure \ref{HDTD}. Our final sample contains 20,089 halo stars, and Figure \ref{sample} shows their velocity distribution in spherical coordinates system.

\section{Substructures detection}
\label{analysis}

\par We now search for local stellar halo substructures in this sample of stars. Models predict that if the halo was built via accretion, the stellar halo in the Solar neighborhood should contain 300-500 streams originating mostly from a handful of massive progenitors \citep{Helmi99a}. The number of streams found in the stellar halo will aid in understanding the role of the past accretion events in the formation of the Milky Way.

\par We distinguish between two uses of the word ``stream" in the literature. One type is dynamical streams, often called ``moving groups," which are groups of stars trapped in a small region of phase space by dynamical resonances. The types stars in such streams, which are often associated with disk resonances, do not depend on their origin, age or type \citep{Dehnen98, Famaey05}. Dynamical streams tend to be clustered in ($U$,$V$) space \citep{Antoja08,Zhao09,Zhao14}. \citet{Klement08} and \citet{Zhao14} detected strams in $V$ and $\sqrt{U^2 + 2V^2}$ space. The other type of stream is tidal streams, which originate from the tidal disruption of accreted satellite galaxies or star clusters. Although the stars in a tidal stream are usually confined to a fraction of the volume of the stellar halo, in a small volume these tidal streams will manifest themselves as a group of stars moving with similar velocities, and in this case tidal streams could be identified as ``moving groups" of stars. \citet{Helmi99b,Helmi17} used a method of searching for streams in ``Integrals of Motion" space, defined by a star's energy, $E$, and two components of the angular momentum. Although this technique can in principle be used to link stars in different volumes to the same infall event, it is often used to detect halo streams in the local solar neighborhood \citep{Kepley07,Helmi00,Helmi17}, because that is the only region for which we presently have 6D phase space information. In this paper, we follow a similar approach to identify local substructure that could correspond to accreted satellites.

\begin{figure*}[h]
	\centering
	\includegraphics[width=0.85\textwidth]{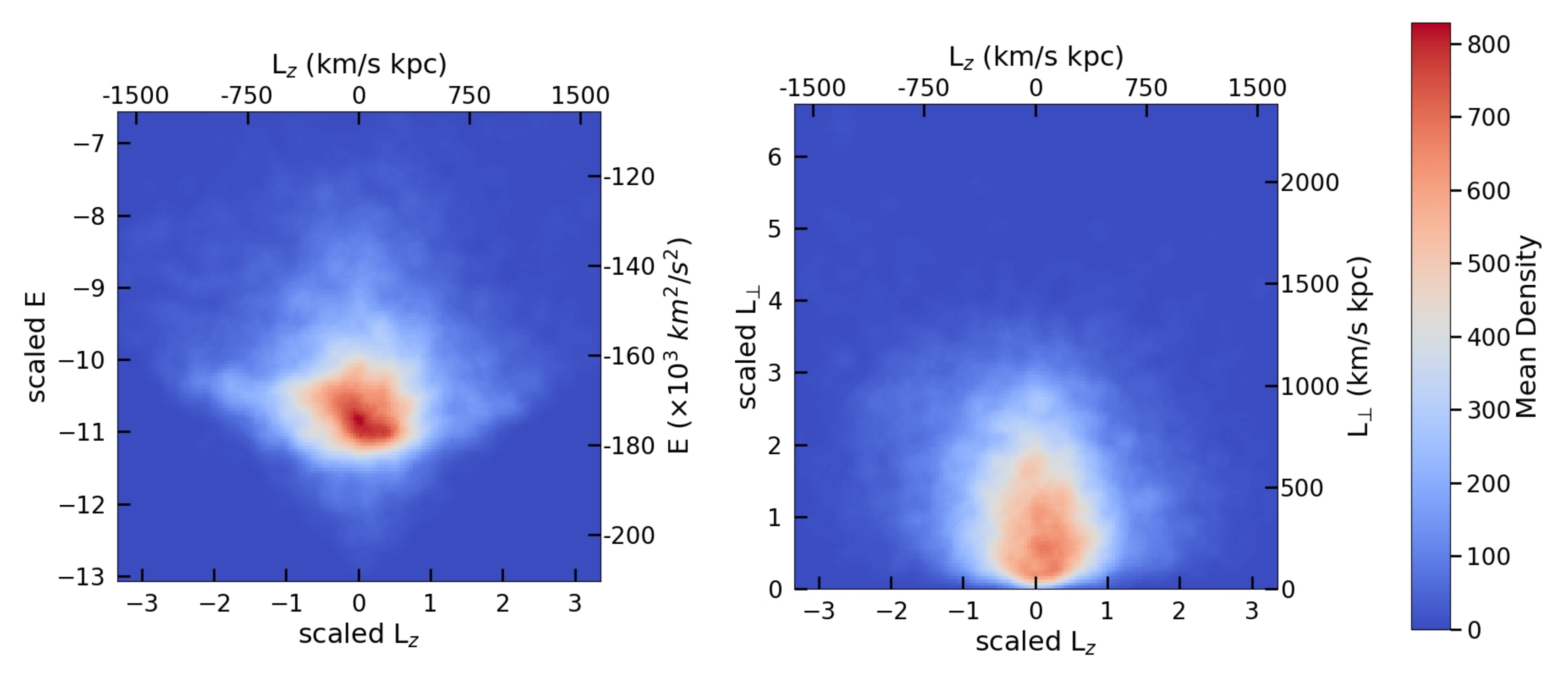}
	\caption{ Projection of three-dimension kernel density distribution of the halo sample stars in $E$ vs $L_z$ space (left panel), and in $L_{\bot}$ vs $L_z$ space (right panel). The color corresponds to their mean density on the third axis ($L_{\bot}$ for the left panel and $E$ for the right panel).}
	\label{kde}
\end{figure*}

\begin{figure*}[h]
	\centering
	\includegraphics[width=0.85\textwidth]{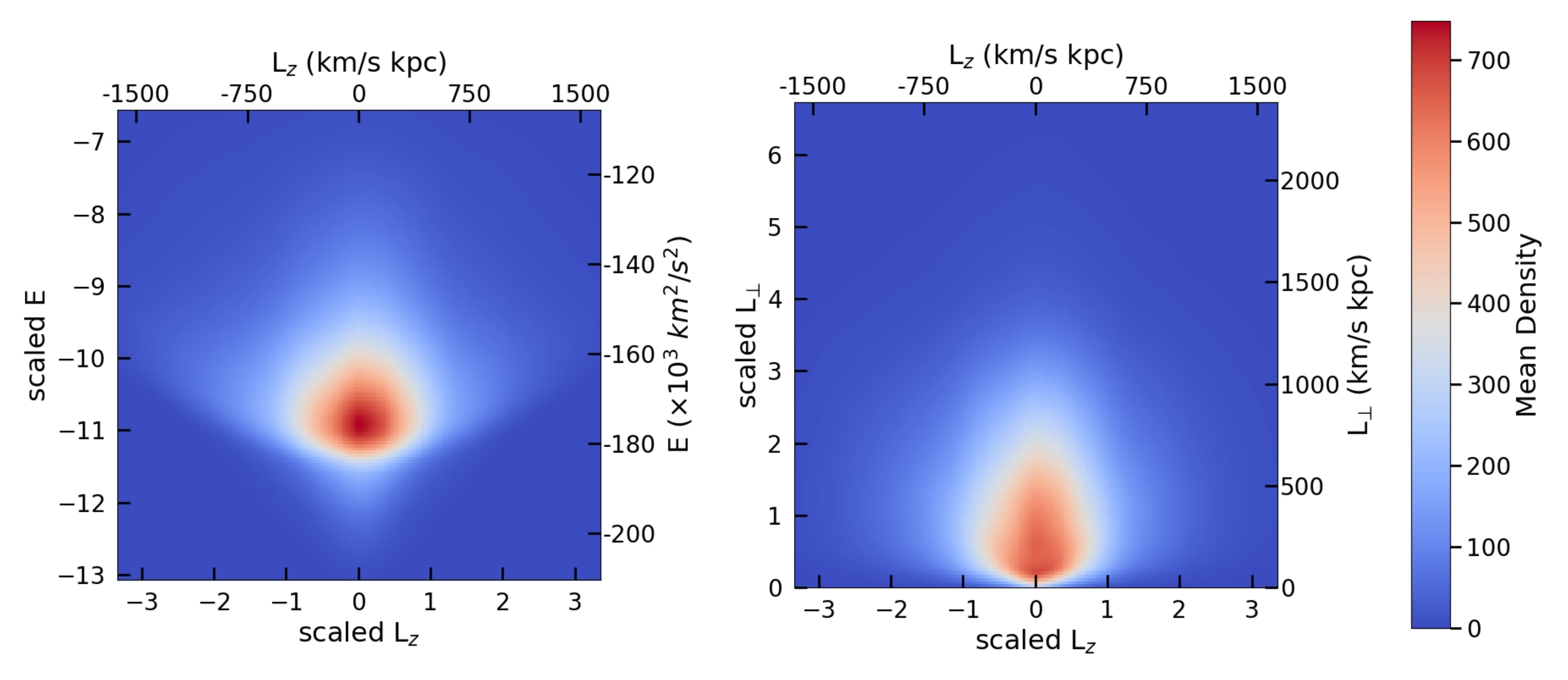}
	\caption{As in Figure \ref{kde}, but now we use average density of all 5000 randomized realizations of the data as the real data.}
	\label{lambda}
\end{figure*}

\begin{figure*}[b]
	\centering
	\includegraphics[width=0.85\textwidth]{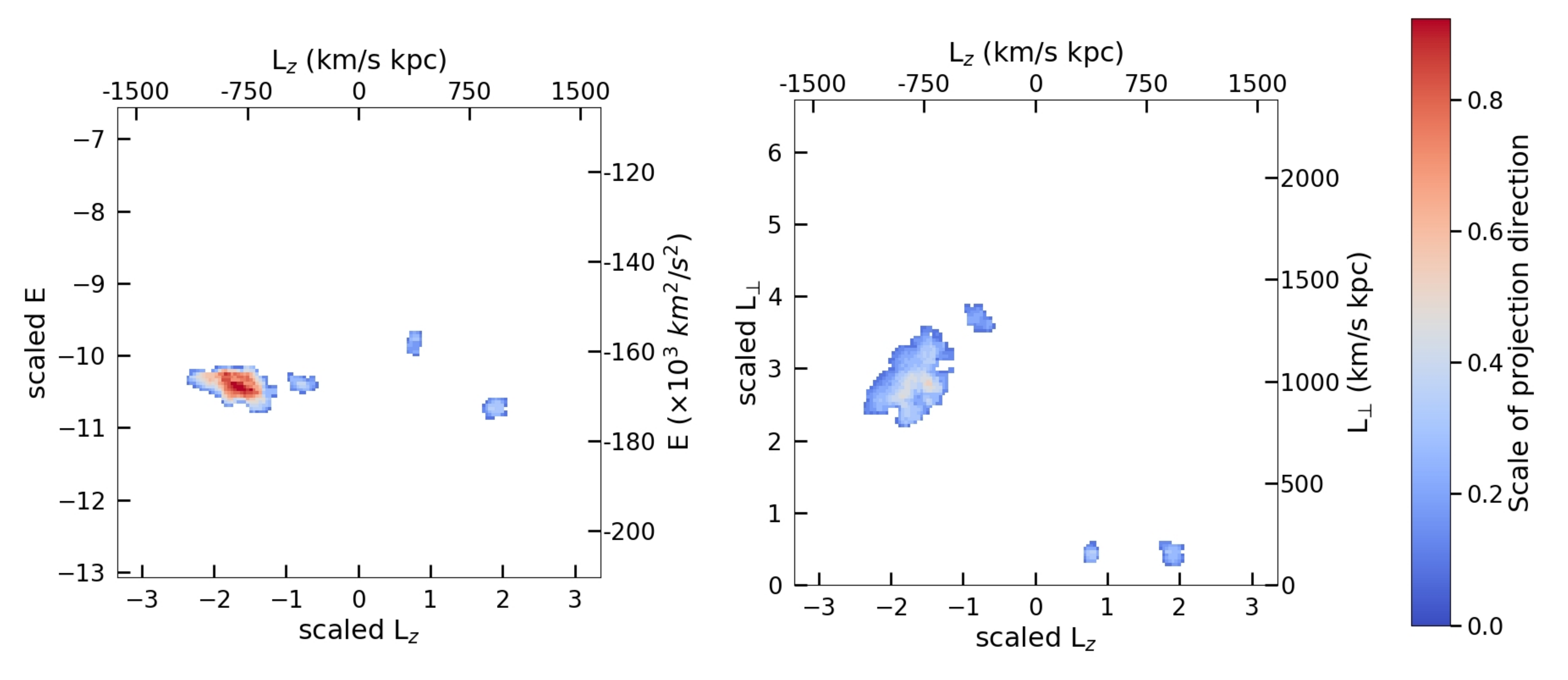}
	\caption{To find the candidate of overdensities, we use the Cumulative Distribution Function of Poisson distribution to estimate the significant of pixels. Only the pixels having a significant $> 3.5\sigma$ and $\lambda >2$ are marked. The color corresponds to their scale length of the third axis.}
	\label{real}
\end{figure*}

\subsection{Distribution of stars in Integrals-of-Motion space}

\par For each of the stars in our halo sample, we compute the energy and angular momentum by adopting a Galaxy potential model which is provided by \citet{McMillan17}. Their model includes four components: the cold gas discs near the Galactic plane, the thin and thick stellar discs, a bulge and a dark-matter halo. Figure \ref{ELL} shows the distribution in energy versus $z$-angular momentum, $L_z$, in the upper panel, and the $L_{\bot} = \sqrt{L^2_x+ L^2_y}$ versus $L_z$ on the lower panel for our sample stars. Note that the $L_z$ of retrograde stars is positive. \citet{Helmi17} found slightly retrograde at $L_z \sim -500$ km s$^{-1}$ kpc and a very retrograde orbits with $L_z < -1000$ km s$^{-1}$ kpc and they concluded that the retrograde rotation was not an artifact of the errors or even an incorrect value for the circular velocity of the Local Standard of Rest. But in this work, the retrograde rotation is not clear.

\subsection{Substructures in the phase-space}

\par In order to constrain our sample in populated region of Figure \ref{ELL},  we select those stars within $-1600 < L_z < 1600$ km s$^{-1}$ kpc, $-2.10 \times 10^5 < E < -1.05 \times 10^5\ \mathrm{km^2\ s^{-2}}$, and $L_{\bot} < 2400$ km s$^{-1}$ kpc. This region is also shown in Figure \ref{ELL}. We apply the $\textsc{sci-kit learn}$ implementation of a non-parametric density estimator that uses an ``Tophat'' kernel to determine the density field of the stars in $E - L_z - L_{\bot}$ space.  We have scaled the data to unit standard deviation to obtain optimal results from the kernel density estimator. The cross-validation method \citep[e.g.][]{Weiss91}, also implemented in $\textsc{sci-kit learn}$, is used to search the optimal bandwidth for the kernel density estimator, and found it to be 0.196.  Then we estimate the density which we call $kde_{\mathrm{real}}$. The result is shown in Figure \ref{kde}. The left panel represents projection of three-dimension kernel density distribution of the halo sample stars in $E$ versus $L_z$ space, and the right is in $L_{\bot}$ versus $L_z$ space. The color corresponds to their mean density on the third axis.

\begin{figure*}[htb]
	\centering
	\includegraphics[width=1.0\textwidth]{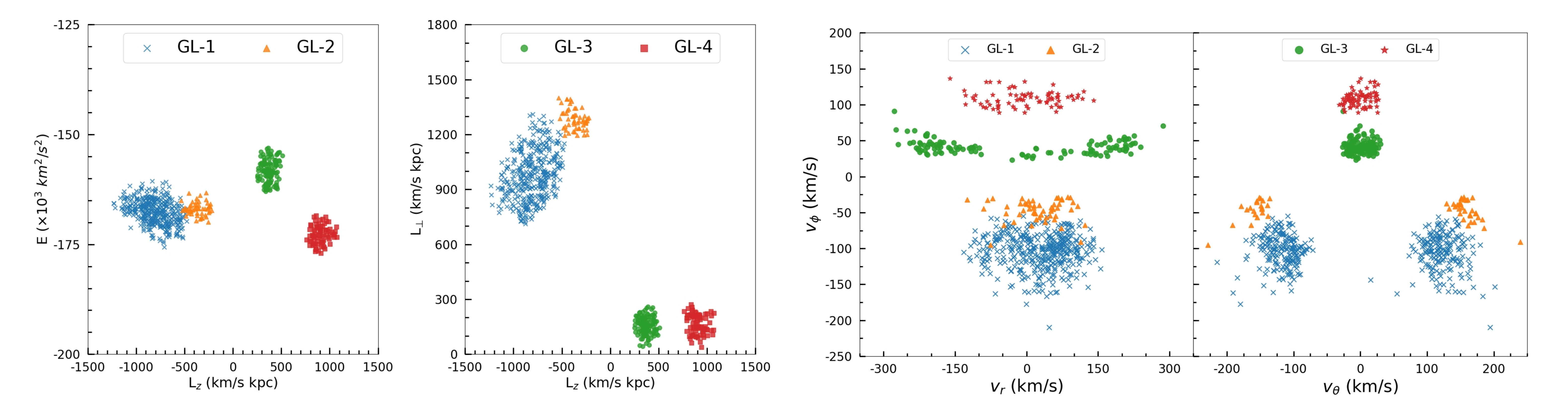}
	\caption{Distribution in $E - L_z - L_{\bot}$ (left panel) and spherical velocities (right panel) space, for the stars comprising the identified substructures.}
	\label{struct}
\end{figure*}

\par To determine the probability that any overdensities in the real data could happen by chance, we reshuffle the velocity components to create 5,000 randomized datasets. Then we recompute their distribution in $E - L_z - L_{\bot}$ space and use the kernel density estimator with same parameters to get the density.  The average density of these randomized datasets is shown in Figure \ref{lambda}. We assume that the part of overdensity is still overdensity, so we use pixels as probes to find the region that might be the overdensity. For each pixel in $E - L_z - L_{\bot}$ space, we can derive the number of stars in a spherical region with a radius of bandwidth from the density of this pixel. We call that number $N_{\mathrm{real}}$ for the real data and $N_{\mathrm{rand}}$ for the randomized datasets. Using 5000 randomized datasets, we can get the mean of $N_{\mathrm{rand}}$ and we call it $\lambda$. We use the Cumulative Distribution Function of Poisson distribution,
\begin{equation}
P(X<N) = \sum_{k=0}^{N-1} \dfrac{\lambda^k}{k!} \,e^{-\lambda}
\end{equation}
to estimate the significant of each pixel in real data. Then we select pixels with significant $> 3.5 \sigma$ and $\lambda >2$. We use $\textsc{sci-kit image}$ \citep{Walt14} implementation to remove small bright spots (i.e. the pixel we selected) and connect small dark cracks. We also remove the size of overdensities smaller than 130 pixels. Result of this procedure is shown in Figure \ref{real}. Adjacent pixels are considered to belong to the same overdensity. When we perform a similar analysis using 100 random datasets as the real data, we find that only one of them appear one overdensity.

\begin{figure*}[htb]
	\centering
	\includegraphics[width=1.0\textwidth]{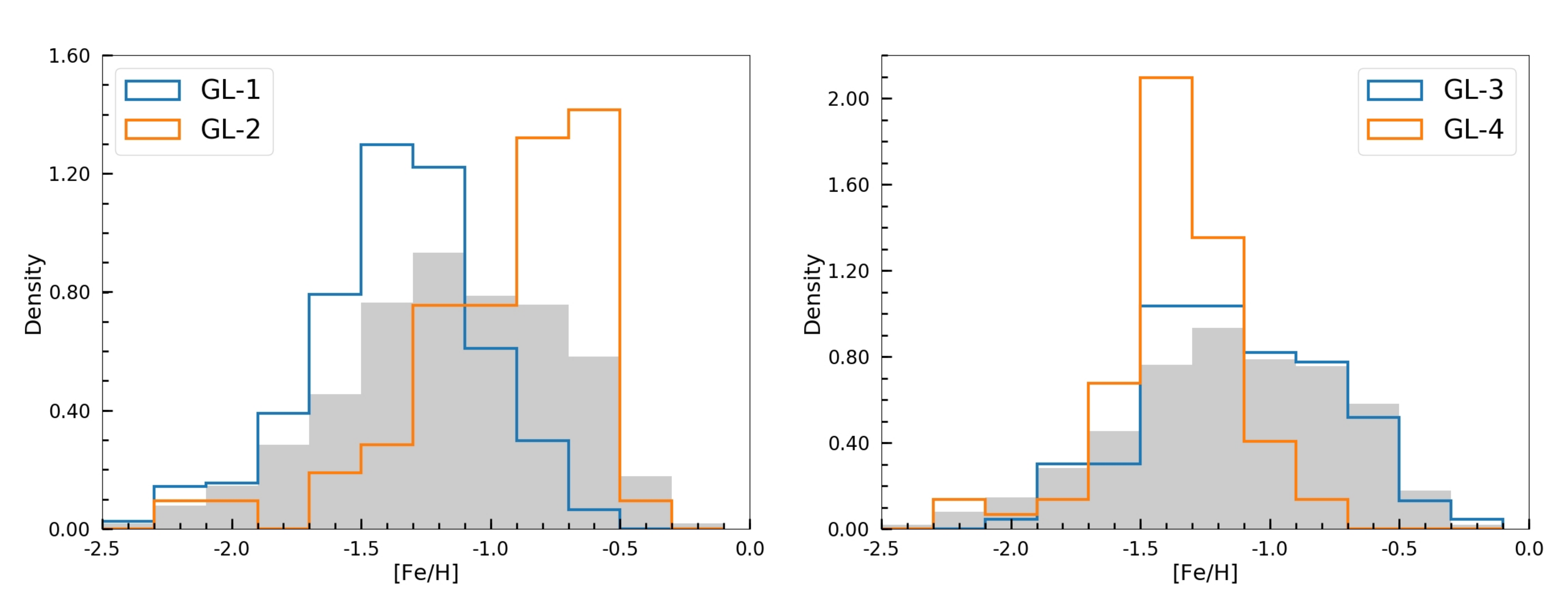}
	\caption{\textbf{Metallicity distribution for the member stars of comprising the identified substructures. Halo sample is shown as gray histogram.}}
	\label{feh}
\end{figure*}

\par Finally, we identify 4 candidates of overdensities in ``Integrals of Motion" space and we extract member stars from the spherical region corresponding to the pixel. We also use the Cumulative Distribution Function of Poisson distribution to estimate the significant of these overdensities.
All of the overdensities are identified substructures in this study, and we label them GL-1 through 4. Distribution of member stars in $E - L_z - L_{\bot}$ space are shown in Figure \ref{struct}. In the Appendix we have tabulated the information of these member stars which belong to substructures. GL-1 has 385 stars, so it not included in the appendix.

\par Figure \ref{struct} also shows the velocity distribution of the member stars in the 4 substructures. Some structures are clustered in $v_{\phi} - v_{\theta}$ space, but separated in $v_{\phi} - v_{r}$ space, e.g. GL-3. Or conversely, they are separated in $v_{\phi} - v_{\theta}$ space, but clustered in $v_{\phi} - v_{r}$ space, e.g. GL-1. This separation may be related to multiple accumulation history. Figure \ref{feh} shows the metallicity distribution for these member stars. As shown in left panel of Figure \ref{struct},  GL-1 and GL-2 are very close in $E - L_z - L_{\bot}$ space, but at their junction, the number of stars is small and the significant of this region is not high enough. In Figure \ref{feh}, we also can see the difference of these two substructures in [Fe/H], so we consider they are different structures.

\subsection{New and old substructures}
\label{new and old}

\par We compared our substructures with other reported substructures \citep[e.g.][]{Klement09,Re Fiorentin15} and found a overlap. The substructure GL-1 identified in our sample and ``N2" stream \citep{Zhao15} have similar [Fe/H], $V_{az}$ and $V_{\Delta E}$. The distributions of the part of GL-1 in our sample are also similar to VelHel-7 of \cite{Helmi17}. It indicate that the streams are probably related to each other, and could originate from the same population. It is not surprising that only one group are found in common with these surveys that using RAVE data, since there is little overlap between the RAVE and the LAMOST dataset. \citet{Li17} presented the candidate members of the Pal 5, GD-1, Cetus Polar and Orphan tidal stellar streams found in LAMOST DR3, SDSS DR9 and APOGEE catalogs. But we found no overlap between their stellar streams and our substructures. \citet{Liang17} applied the wavelet transform method to identify 16 significant overdensities in velocity space. Due to the selection effects of sample stars, e.g. different halo samples or the space used to identify the substructure, those star streams found in their surveys are not included completely in our cross-matched data set. In summary, the significant, mean energy, $L_z$ and $L_{\bot}$ of 4 structures are shown in the Table \ref{mean}. A catalog of 3 substructures' members (except the largest structure) are given in the Appendix.

\begin{center}
	\begin{longtable}{ccccc}
		\label{mean}\\
		\caption{The center of the overdensities in our sample}\\
		\hline
		\hline
		Notation &  significant  &  E  &    $L_z$    &    $L_{\bot}$    \\
		& & ($\times 10^3\ \mathrm{km^2\ s^{-2}}$)   &  (km s$^{-1}$ kpc) & (km s$^{-1}$ kpc) \\
		\hline
		    GL-1 & $>10\,\sigma$  & -167.6 &  -808.8 &  977.8 \\
		    GL-2 & $5.17\,\sigma$ & -166.9 &  -367.8 & 1284.0 \\
		    GL-3 & $4.18\,\sigma$ & -158.3 &   370.8 &  156.4 \\
		    GL-4 & $5.10\,\sigma$ & -172.8 &   903.9 &  165.8 \\
		\hline
	\end{longtable}
\end{center}

\section{Conclusions}
\label{conclusion}

\par By cross-matching the Gaia DR2 and LAMOST DR5 data, we obtain a sample of stars with full phase-space information and identified a subset sample halo stars kinematically and chemically. Our final halo sample consists of 20,089 stars.  We determine the distribution of the sample halo stars in ``Integrals-of-Motion" space, defined by two components of the angular momentum, $L_{\bot}$ and $L_z$, and by energy. We find some overdensities in this space. To remove the contamination,  we estimate the statistical significance for these overdensities and identify 4 substructures. Three of the 4 significant structures are not reported in previous works. One structure may correspond to the ``N2" stream of \cite{Zhao15} and VelHel-7 of \cite{Helmi17}. The identified structures may be the debris from satellite accretion events, their dynamical and chemical properties can aid in understanding the history of the Milky Way. 

\section{Acknowledgements}

\par We thank especially the referee for insightful comments and suggestions, which have improved the paper significantly. This work has made use of data from the European Space Agency (ESA) mission {\it Gaia} (\url{https://www.cosmos.esa.int/gaia}), processed by the {\it Gaia} Data Processing and Analysis Consortium (DPAC, \url{https://www.cosmos.esa.int/web/gaia/dpac/consortium}). Funding for the DPAC has been provided by national institutions, in particular the institutions participating in the {\it Gaia} Multilateral Agreement.

\par This work was supported by joint funding for Astronomy by the National Natural Science Foundation of China and the Chinese Academy of Science, under Grants U1231113. This work was also by supported by the Special funds of cooperation between the Institute and the University of the Chinese Academy of Sciences, and China Scholarship Council (CSC). HJN acknowledges funding from NSF grant AST-1615688. Funding for SDSS-III has been provided by the Alfred P. Sloan Foundation, the Participating Institutions, the National Science Foundation, and the U.S. Department of Energy Office of Science. This project was developed in part at the 2016 NYC Gaia Sprint, hosted by the Center for Computational Astrophysics at the Simons Foundation in New York City.  The Guoshoujing Telescope (LAMOST) is a National Major Scientific Project has been provided by the National Development and Reform Commission. LAMOST is operated and managed by the National Astronomical Observatories, Chinese Academy of Sciences.

\newpage
\appendix  
\renewcommand{\appendixname}{Appendix~\Alph{section}}  
\section{Member stars of the new identified substructures}
\centering
\par Unique source identifiers in Gaia and LAMOST, positions, metallicities, space velocities and z-angular momentum for the member stars.

\begin{center}
	\begin{longtable}{cccccccccccc}
\caption{GL-2, 53 stars}\\
\hline
\hline
source-id &      obsid &       l &      b &  [Fe/H] &     x &    y &     z &      U &      V &      W &     $L_z$     \\
		  &            & (deg)   & (deg) 　& (dex)   & (kpc) & (kpc)& (kpc) & (km s$^{-1}$) & (km s$^{-1}$) & (km s$^{-1}$) & (km s$^{-1}$ kpc) \\
\hline\\
 3662032469693724800 &  572211196 &  331.38 &  59.02 &   -1.31 &  6.0 &  1.2 &  4.2 &   10.8 &   48.8 &  193.2 & -278.3 \\
2503115396900046720 &  209808189 &  166.72 & -51.47 &   -0.64 &  8.6 & -0.1 & -0.5 &  -64.2 &   31.5 & -154.8 & -264.6 \\
1231233915353158528 &  120207106 &    1.90 &  69.06 &   -0.83 &  7.9 & -0.0 &  0.9 &  -61.4 &   57.0 & -169.9 & -447.1 \\
1285189081231978752 &  233011085 &   49.30 &  69.95 &   -1.20 &  7.9 & -0.4 &  1.3 &  -18.9 &   47.7 & -166.3 & -369.4 \\
1286611711842881024 &  227006099 &   53.82 &  64.70 &   -0.51 &  7.8 & -0.6 &  1.5 &   52.9 &   28.9 & -171.9 & -254.6 \\
1484880665082822528 &  422712018 &   71.79 &  68.72 &   -1.63 &  8.1 & -0.3 &  0.9 &  -44.7 &   45.5 &  170.8 & -354.3 \\
1545642735656223744 &  448503191 &  138.17 &  67.81 &   -0.89 &  8.5 & -0.3 &  1.1 &  -17.4 &   38.4 & -146.4 & -322.9 \\
3886582740137269248 &  275101172 &  225.58 &  58.38 &   -0.55 &  8.9 &  0.7 &  1.6 &  -94.6 &   30.8 & -125.5 & -338.4 \\
1247776480110134272 &  450215050 &    0.84 &  75.26 &   -0.94 &  7.7 & -0.0 &  2.1 &  -51.5 &   68.2 & -153.2 & -522.1 \\
2653377439090572928 &  362206040 &   70.94 & -50.33 &   -1.13 &  7.7 & -1.4 & -1.8 & -120.9 &   58.1 &  130.4 & -278.9 \\
4368228034935117184 &  457401037 &   19.41 &  21.30 &   -1.17 &  4.6 & -1.3 &  1.5 & -150.1 &  135.7 & -194.0 & -432.3 \\
4386425158532864384 &  327606021 &   22.57 &  28.14 &   -1.44 &  4.6 & -1.5 &  2.1 &  182.3 &   40.7 &  179.2 & -460.1 \\
4408486687546910464 &  552612050 &   13.43 &  33.77 &   -0.43 &  6.8 & -0.3 &  1.0 &  -89.3 &   71.8 &  207.5 & -455.8 \\
4446871703622166144 &  240813081 &   27.43 &  33.82 &   -1.25 &  6.7 & -0.8 &  1.1 &  -70.9 &   49.4 &  193.7 & -277.8 \\
4451604345266684160 &  233105084 &   19.65 &  40.92 &   -0.67 &  7.4 & -0.3 &  0.7 &  109.0 &   28.0 & -172.0 & -238.2 \\
2120723115527908096 &  155706110 &   77.60 &  22.84 &   -2.24 &  7.4 & -3.6 &  1.6 &  -19.3 &   42.1 &  161.4 & -241.4 \\
291660753646824320 &  267302070 &  136.30 & -37.55 &   -0.64 &  8.6 & -0.4 & -0.4 &  -65.1 &   57.8 &  135.3 & -472.1 \\
2747259239428992000 &  182916121 &  110.41 & -56.70 &   -0.83 &  8.4 & -0.6 & -0.9 &  108.3 &   37.1 & -137.5 & -375.2 \\
2801278108262214912 &  285308088 &  121.43 & -42.88 &   -0.56 &  8.7 & -0.8 & -0.8 &  -26.4 &   43.1 &  146.5 & -353.8 \\
2835956705002706176 &  353313048 &   89.90 & -33.05 &   -0.55 &  8.2 & -1.1 & -0.7 &  -68.1 &   38.0 & -161.8 & -234.3 \\
2864046890629860480 &   81903062 &  115.83 & -28.38 &   -0.59 &  8.7 & -1.0 & -0.6 &  -18.1 &   52.4 &  145.6 & -436.7 \\
3390618402335173632 &  374410052 &  188.43 & -12.29 &   -0.58 &  8.8 &  0.1 & -0.1 &   69.1 &   31.2 &  136.8 & -269.2 \\
3719566167963752576 &  346103184 &  332.39 &  68.58 &   -1.30 &  8.0 &  0.1 &  0.5 &   15.7 &   63.2 & -172.0 & -505.8 \\
3729537260959070848 &  422111089 &  314.63 &  69.49 &   -1.27 &  7.9 &  0.3 &  1.2 &  -45.0 &   48.9 & -168.2 & -399.7 \\
3823022476900895104 &  415910187 &  242.85 &  37.42 &   -0.91 &  8.6 &  0.8 &  0.7 &    5.6 &   34.4 &  155.0 & -292.1 \\
3858667372562954880 &  215105115 &  242.23 &  51.56 &   -0.85 &  8.5 &  0.6 &  0.9 &  -15.4 &   39.0 &  141.2 & -340.7 \\
3879706050819568384 &  484603106 &  224.86 &  45.27 &   -0.82 &  8.8 &  0.6 &  0.8 &   -3.5 &   43.4 &  148.0 & -382.6 \\
98563009748281472 &  378512245 &  141.66 & -37.88 &   -0.65 &  8.7 & -0.4 & -0.5 &  -23.0 &   49.1 & -158.1 & -419.2 \\
586365363499961728 &  223808024 &  225.52 &  37.43 &   -0.71 &  8.4 &  0.2 &  0.3 &  -68.5 &   27.9 & -148.8 & -250.7 \\
598769607008444672 &  420802068 &  216.12 &  29.88 &   -0.52 &  8.9 &  0.5 &  0.6 &   30.2 &   43.7 & -144.3 & -374.6 \\
659641078984352384 &  400306238 &  207.10 &  31.27 &   -0.93 &  8.5 &  0.2 &  0.2 &  -23.0 &   56.4 &  155.5 & -483.0 \\
726491312936718208 &  139813093 &  208.51 &  53.61 &   -1.17 &  8.6 &  0.2 &  0.7 &   -2.0 &   33.1 &  147.1 & -286.3 \\
760791200935729792 &  450115068 &  177.30 &  69.55 &   -1.70 &  8.4 & -0.0 &  0.5 &   33.9 &   51.3 & -151.5 & -431.0 \\
1246720742788984320 &  558005076 &    9.84 &  73.36 &   -0.57 &  7.5 & -0.1 &  2.3 & -114.3 &   33.9 & -127.9 & -242.8 \\
1278545935057357824 &  350215109 &   53.58 &  57.10 &   -1.26 &  7.1 & -1.5 &  3.0 &  -94.1 &   75.8 & -133.9 & -392.6 \\
1336301837755473280 &  462608072 &   60.77 &  31.54 &   -2.06 &  6.5 & -3.0 &  2.1 &  111.8 &   23.5 &  143.8 & -486.0 \\
1491909293163433344 &  575415144 &   76.31 &  67.28 &   -0.95 &  8.1 & -0.3 &  0.8 &  -43.1 &   55.0 & -160.3 & -433.4 \\
1525146464518953088 &  210910197 &   98.93 &  74.21 &   -1.21 &  8.2 & -0.1 &  0.5 &  -65.4 &   39.8 & -145.8 & -318.3 \\
1590711339315440512 &  418016063 &   83.85 &  59.06 &   -0.51 &  8.1 & -0.7 &  1.1 &  -31.1 &   47.9 &  153.7 & -368.8 \\
1731977452245467520 &  370212003 &   53.87 & -28.37 &   -0.70 &  6.5 & -2.3 & -1.5 &   -5.8 &   78.1 & -196.2 & -496.4 \\
1739454062594964352 &  469604140 &   60.53 & -29.61 &   -0.81 &  7.9 & -0.5 & -0.3 &  -54.4 &   51.5 &  157.4 & -378.4 \\
1864801343109259904 &  475212086 &   75.34 &  -9.32 &   -0.84 &  7.3 & -3.5 & -0.6 &   16.2 &   49.0 & -172.9 & -413.6 \\
1304558662184239232 &  225006050 &   47.15 &  39.74 &   -0.74 &  7.7 & -0.6 &  0.7 & -103.6 &   36.9 & -158.7 & -220.9 \\
2511535418290211328 &  384306035 &  154.75 & -56.42 &   -1.01 &  8.4 & -0.1 & -0.4 &    9.3 &   52.6 &  158.2 & -444.8 \\
912354579562114816 &  181801156 &  181.48 &  37.88 &   -0.78 &  8.6 &  0.0 &  0.3 &  -57.6 &   42.7 & -150.0 & -366.8 \\
4009658429618705152 &  332707210 &  209.88 &  82.20 &   -0.72 &  8.6 &  0.2 &  3.0 &  -50.5 &   43.3 & -123.8 & -380.4 \\
4006785027777824384 &  332705203 &  207.35 &  80.19 &   -1.03 &  8.5 &  0.1 &  1.7 &  -27.0 &   36.7 & -147.3 & -314.3 \\
1740820648172737280 &  377801243 &   63.00 & -31.39 &   -0.76 &  7.4 & -1.6 & -1.1 &   19.0 &   57.0 & -183.5 & -451.1 \\
612386543123006336 &  554316077 &  209.21 &  36.83 &   -0.54 &  9.4 &  0.7 &  1.1 &   -3.8 &   37.3 & -130.1 & -354.5 \\
3693711839232555008 &  208504143 &  293.01 &  60.12 &   -0.63 &  7.8 &  0.8 &  1.6 &  -94.4 &   49.6 & -139.3 & -468.6 \\
604279569012986752 &   94015151 &  217.25 &  35.20 &   -0.96 &  8.4 &  0.2 &  0.2 &   13.1 &   30.5 &  149.1 & -255.1 \\
2746111727246997376 &  505912064 &   99.03 & -53.83 &   -0.96 &  8.3 & -0.9 & -1.2 &   18.4 &   37.4 & -144.9 & -328.3 \\
133443366875832832 &  388104051 &  148.95 & -25.45 &   -0.72 &  8.5 & -0.2 & -0.1 &  -35.5 &   63.9 & -166.0 & -535.4 \\
\hline
\end{longtable}
\end{center}

\newpage
\begin{center}
	\begin{longtable}{cccccccccccc}
\caption{GL-3, 116 stars}\\
\hline
\hline
source-id &      obsid &       l &      b &  [Fe/H] &     x &    y &     z &      U &      V &      W &     $L_z$     \\
&            & (deg)   & (deg) 　& (dex)   & (kpc) & (kpc)& (kpc) & (km s$^{-1}$) & (km s$^{-1}$) & (km s$^{-1}$) & (km s$^{-1}$ kpc) \\
\hline
  776435773930211328 &  142613029 &  178.19 &  62.15 &   -0.68 &  11.0 & -0.1 &  5.2 &   30.4 &  -23.3 &   -7.4 &  252.6 \\
2521025925920389504 &  266810185 &  154.78 & -50.88 &   -1.19 &   8.9 & -0.3 & -0.9 &  187.1 &  -47.2 &   -4.0 &  360.2 \\
3955108049991504512 &  101501160 &  288.27 &  84.94 &   -1.73 &   8.2 &  0.0 &  0.5 &  215.9 &  -42.5 &    9.4 &  357.5 \\
599099323058438016 &  313815016 &  216.50 &  24.85 &   -0.91 &  10.1 &  1.4 &  1.1 &  152.6 &  -11.5 &  -18.8 &  336.5 \\
605566719171720960 &  502101137 &  214.54 &  33.94 &   -1.42 &   8.9 &  0.5 &  0.6 &  193.7 &  -28.2 &  -26.5 &  344.4 \\
711037268988125312 &  130104169 &  189.83 &  35.60 &   -1.35 &  10.5 &  0.4 &  1.7 &  139.2 &  -35.8 &  -30.4 &  431.7 \\
786574954766550784 &  146707169 &  148.02 &  66.27 &   -1.25 &   9.1 & -0.6 &  2.5 &  142.2 &  -47.0 &  -19.2 &  346.8 \\
790728561803079168 &  314703069 &  149.31 &  63.35 &   -1.49 &   8.6 & -0.3 &  1.0 & -209.8 &  -44.3 &   -3.7 &  437.5 \\
828495278684213888 &  293914160 &  162.09 &  48.50 &   -0.92 &  10.9 & -0.9 &  3.2 &  105.3 &  -48.7 &  -41.3 &  438.4 \\
829090492436805504 &  343413114 &  173.15 &  57.62 &   -1.08 &  10.6 & -0.3 &  3.8 &   97.7 &  -42.7 &  -17.1 &  423.8 \\
166891403944921344 &  279604032 &  165.36 & -17.94 &   -0.84 &  10.8 & -0.7 & -0.8 &  147.9 &  -56.5 &   24.5 &  509.9 \\
1533388334960428672 &  132809043 &  143.96 &  75.07 &   -0.80 &   9.1 & -0.6 &  4.1 &  105.7 &  -42.9 &  -42.1 &  321.6 \\
1542725937461194752 &  319209158 &  126.81 &  70.32 &   -0.38 &   8.8 & -0.8 &  2.9 & -163.5 &  -25.6 &   62.1 &  358.4 \\
1545461281877377664 &  145105114 &  136.24 &  68.61 &   -1.16 &   8.5 & -0.2 &  0.9 & -219.0 &  -40.1 &   48.6 &  391.0 \\
1554494044772634368 &  149701075 &  111.65 &  68.51 &   -0.88 &   8.7 & -1.3 &  3.7 & -127.2 &  -30.3 &   63.7 &  435.4 \\
1597431554384379264 &  153713095 &   85.61 &  47.67 &   -0.99 &   8.1 & -1.1 &  1.3 &  215.2 &  -68.1 &  -53.8 &  309.4 \\
4471882393403103232 &  347705157 &   33.08 &  12.77 &   -0.96 &   4.4 & -2.5 &  1.1 &  196.9 & -215.1 &  -31.1 &  459.1 \\
691295881309324800 &  107208219 &  198.03 &  36.56 &   -1.55 &  11.5 &  1.1 &  2.6 &  -19.7 &  -40.9 &  -11.2 &  447.7 \\
649320959766434560 &  337414159 &  211.93 &  25.54 &   -0.82 &  11.1 &  1.8 &  1.7 & -124.3 &  -56.2 &   30.7 &  398.3 \\
2825552026469603968 &  253004025 &   97.20 & -38.41 &   -1.34 &   8.6 & -2.8 & -2.2 &  133.8 &  -84.8 &   43.6 &  350.3 \\
4391534863945071616 &  449915052 &   23.88 &  26.25 &   -1.14 &   3.8 & -1.9 &  2.4 & -253.9 &   51.0 &  140.8 &  300.6 \\
4422824525328500224 &  564606181 &    7.49 &  41.40 &   -0.72 &   6.3 & -0.2 &  1.7 &  252.1 &  -54.9 &  -93.5 &  283.2 \\
4427480132438480384 &  135310139 &    9.04 &  44.96 &   -1.09 &   7.3 & -0.1 &  0.9 & -237.4 &  -36.8 &   33.4 &  302.9 \\
4453040131355904768 &  236616049 &   22.02 &  39.17 &   -1.30 &   6.3 & -0.8 &  1.7 &  229.0 &  -91.2 &  -79.8 &  406.1 \\
4543000115454359808 &  342805021 &   38.60 &  24.33 &   -1.29 &   3.8 & -3.5 &  2.5 &  140.2 & -214.3 & -116.5 &  337.9 \\
2130702527879104512 &  354401051 &   77.72 &  17.35 &   -1.44 &   7.0 & -5.3 &  1.7 &  117.9 & -145.4 &  -53.5 &  394.6 \\
2145762126325789440 &  153801182 &   81.48 &  24.30 &   -1.73 &   7.3 & -6.0 &  2.8 &   77.7 & -113.1 &  -26.2 &  361.2 \\
2149435629033487744 &  155308036 &   81.63 &  27.82 &   -1.87 &   7.5 & -4.6 &  2.5 &  130.1 & -127.9 &  -36.3 &  365.4 \\
2150266000831014400 &  155311112 &   84.07 &  28.90 &   -0.82 &   8.0 & -2.2 &  1.2 &  184.2 &  -97.3 &  -32.9 &  372.6 \\
2704791427439158912 &  247805073 &   71.19 & -45.24 &   -1.05 &   7.7 & -1.5 & -1.6 &  188.4 &  -87.1 &   26.4 &  377.8 \\
2733432605831478144 &  174808087 &   81.93 & -36.55 &   -0.55 &   7.9 & -2.0 & -1.5 &  183.0 &  -89.5 &   64.0 &  334.8 \\
2735977288054696832 &  480805041 &   79.68 & -35.88 &   -1.37 &   8.1 & -0.5 & -0.4 & -210.5 &  -30.3 &    8.6 &  360.1 \\
2805435082553510400 &   10805107 &  125.24 & -36.66 &   -0.42 &  10.7 & -3.5 & -3.2 &   84.9 &  -72.9 &   31.1 &  479.8 \\
2807439668344927104 &  194514093 &  117.85 & -37.82 &   -1.44 &   9.9 & -3.2 & -2.8 &   74.3 &  -56.6 &   41.0 &  321.4 \\
317684059276916608 &  362611053 &  135.59 & -27.40 &   -1.17 &  10.8 & -2.6 & -1.9 &   94.0 &  -58.8 &   20.9 &  394.4 \\
320491597794951936 &  283908194 &  130.96 & -26.80 &   -1.41 &   8.7 & -0.6 & -0.4 & -188.2 &  -27.8 &  -18.4 &  346.8 \\
3064855269056385152 &  196807059 &  230.42 &  18.51 &   -0.54 &  11.1 &  3.5 &  1.6 &   15.9 &  -26.9 &  -15.1 &  354.9 \\
3073650979825727616 &  137205087 &  226.14 &  22.84 &   -1.43 &   9.5 &  1.4 &  0.8 &  166.0 &  -17.0 &  -29.9 &  387.5 \\
3139910578471472000 &  492403159 &  212.18 &   8.94 &   -1.17 &  11.2 &  1.9 &  0.6 & -137.7 &  -55.8 &   22.3 &  366.4 \\
3360380664340716544 &  270704186 &  200.26 &  12.22 &   -0.94 &   9.9 &  0.6 &  0.4 &  181.7 &  -21.3 &   13.9 &  322.8 \\
381409077459596800 &  176615070 &  120.08 & -21.46 &   -1.23 &   8.3 & -0.2 & -0.1 & -219.0 &  -25.2 &   23.1 &  262.2 \\
3632411557722152320 &  414715155 &  326.41 &  54.72 &   -1.61 &   7.8 &  0.3 &  0.7 &  214.0 &  -50.1 &   -5.7 &  448.9 \\
3710972763038252032 &  212911054 &  323.46 &  61.49 &   -1.10 &   6.7 &  1.1 &  3.5 & -163.5 &  -78.4 &   86.0 &  337.6 \\
3734233034603510144 &  342107109 &  315.26 &  72.02 &   -1.20 &   7.4 &  0.8 &  3.5 &  174.9 &  -26.4 &  -64.0 &  334.2 \\
3884500604416157056 &  387510031 &  232.25 &  56.94 &   -1.36 &   8.3 &  0.2 &  0.3 & -203.2 &  -58.6 &   24.2 &  455.9 \\
3885678597981496192 &  331909154 &  228.54 &  56.64 &   -1.34 &   8.8 &  0.7 &  1.4 & -204.8 &  -51.3 &   14.5 &  314.6 \\
3945085589188023424 &  405505032 &  349.01 &  85.25 &   -1.05 &   7.9 &  0.1 &  3.4 &  189.8 &  -35.0 &  -98.6 &  287.7 \\
4018063508817588864 &  507105216 &  209.70 &  74.14 &   -0.92 &   8.3 &  0.1 &  0.4 &  223.1 &  -32.5 &  -22.9 &  280.8 \\
4023554332447594624 &   20503110 &  195.97 &  71.72 &   -0.72 &   8.3 &  0.0 &  0.3 &  225.3 &  -38.5 &  -31.6 &  325.4 \\
586009671488680320 &  223705163 &  225.29 &  36.69 &   -1.45 &   8.5 &  0.3 &  0.4 & -206.4 &  -47.7 &   26.0 &  338.1 \\
686132643423960448 &  398004026 &  204.19 &  36.86 &   -1.08 &  11.4 &  1.4 &  2.6 &  123.8 &  -23.8 &  -13.6 &  448.8 \\
722295477781603200 &  308501029 &  213.17 &  57.78 &   -0.77 &   8.6 &  0.2 &  0.7 & -207.8 &  -58.3 &    6.6 &  449.8 \\
785873466350271360 &  563009040 &  153.66 &  64.26 &   -0.56 &   9.0 & -0.4 &  1.9 & -170.5 &  -20.9 &   15.9 &  255.7 \\
785710463748905600 &  140703021 &  157.44 &  64.19 &   -1.20 &   9.3 & -0.4 &  2.4 & -177.4 &  -31.8 &   45.8 &  375.0 \\
826632431109291648 &  208413191 &  166.90 &  49.67 &   -1.25 &  10.3 & -0.5 &  2.6 & -123.8 &  -26.5 &   31.4 &  333.1 \\
831609885892678016 &  279407199 &  161.50 &  60.22 &   -1.15 &   8.3 & -0.0 &  0.1 &  222.0 &  -46.1 &  -34.2 &  376.1 \\
892040453700593280 &  314412112 &  187.64 &  21.29 &   -0.63 &  11.9 &  0.5 &  1.5 &  -42.8 &  -35.0 &   15.3 &  396.2 \\
927844606749721984 &  318004152 &  177.49 &  32.35 &   -0.62 &   9.3 & -0.0 &  0.7 &  188.9 &  -33.4 &  -38.5 &  300.5 \\
1022930028722544128 &  265809105 &  163.30 &  43.89 &   -0.76 &   9.8 & -0.5 &  1.6 & -150.9 &  -21.9 &   42.5 &  284.8 \\
1024562769129570688 &  528303141 &  158.89 &  42.00 &   -0.74 &   9.1 & -0.4 &  0.9 & -196.0 &  -29.4 &    8.6 &  340.1 \\
1193391920583495552 &  347607079 &   28.35 &  46.27 &   -1.05 &   7.9 & -0.2 &  0.4 & -228.7 &  -42.2 &   -8.1 &  370.0 \\
1252847736974969600 &  316811145 &   21.77 &  69.49 &   -1.15 &   6.8 & -0.6 &  4.0 &  183.2 &  -71.9 &  -93.2 &  388.0 \\
1287943254781596160 &  227015146 &   58.31 &  65.94 &   -0.82 &   7.4 & -1.2 &  3.3 &  183.0 &  -87.8 &  -81.3 &  429.3 \\
1327550721630335872 &  555009234 &   57.17 &  41.86 &   -1.23 &   7.3 & -1.4 &  1.5 &  214.9 & -106.9 &  -40.6 &  477.0 \\
1331866403552107008 &   48814092 &   63.18 &  42.19 &   -1.26 &   6.9 & -2.6 &  2.7 &  193.1 & -124.1 &  -76.9 &  351.4 \\
1361879708033005952 &  144807156 &   72.03 &  33.63 &   -1.04 &   6.9 & -4.2 &  2.9 &  138.0 & -151.7 &  -78.1 &  465.6 \\
1415072534396049280 &  145408225 &   78.44 &  34.75 &   -0.84 &   7.2 & -5.0 &  3.5 & -148.5 &   44.8 &   73.3 &  413.6 \\
1460842866998627840 &  301804060 &   62.86 &  87.59 &   -1.41 &   8.2 & -0.1 &  1.6 &  203.8 &  -61.5 &  -44.7 &  490.3 \\
1483498922564228608 &  419503232 &   69.76 &  71.98 &   -0.48 &   7.9 & -0.9 &  3.0 &  159.8 &  -64.9 &  -68.5 &  368.0 \\
1493366588450862976 &  128409018 &   77.42 &  63.23 &   -1.40 &   7.9 & -1.4 &  2.9 & -183.8 &  -10.5 &   58.9 &  346.5 \\
1514561500437195776 &  298106062 &  143.43 &  83.88 &   -1.20 &   8.3 & -0.1 &  1.3 &  191.7 &  -46.1 &  -42.6 &  367.2 \\
1522631400389533568 &  308708228 &  109.02 &  79.13 &   -1.31 &   8.2 & -0.1 &  0.4 &  192.3 &  -57.0 &  -35.1 &  454.8 \\
1560427043875297536 &  397909218 &  109.92 &  62.89 &   -0.90 &   8.6 & -1.1 &  2.2 & -172.2 &  -34.1 &   47.8 &  475.8 \\
1589977346585245056 &  418004010 &   80.72 &  59.12 &   -1.26 &   7.8 & -2.6 &  4.4 &  133.5 &  -89.9 & -100.7 &  357.3 \\
1600092647401514368 &  446808094 &   89.11 &  52.30 &   -1.41 &   8.2 & -2.6 &  3.4 & -145.0 &   -0.0 &   53.3 &  376.3 \\
1612828153147112448 &  564306079 &   94.30 &  52.24 &   -1.00 &   8.3 & -1.1 &  1.5 &  214.4 &  -76.5 &  -27.0 &  396.4 \\
39069122766561920 &  381902155 &  177.21 & -27.52 &   -1.24 &  10.5 & -0.1 & -1.2 &  112.2 &  -41.5 &    5.4 &  422.1 \\
1760073066712199680 &  371605087 &   62.28 & -21.05 &   -0.56 &   5.8 & -4.5 & -1.9 & -189.5 &   80.6 &  -41.2 &  383.8 \\
1893918575666556416 &  469702033 &   87.17 & -24.00 &   -1.70 &   8.0 & -4.5 & -2.0 &  127.1 & -128.4 &   54.6 &  446.0 \\
217189761654075520 &  472511147 &  160.67 & -16.11 &   -1.72 &  12.8 & -1.6 & -1.4 &   -3.7 &  -27.3 &    7.3 &  355.5 \\
1907573200116557312 &  178415015 &   93.52 & -17.60 &   -0.74 &   8.3 & -1.8 & -0.5 & -210.4 &    3.3 &  -31.3 &  348.7 \\
1939324695044566400 &  182610080 &  111.15 & -13.73 &   -1.55 &  10.2 & -5.3 & -1.4 &  -97.4 &   10.7 &  -12.5 &  401.9 \\
3155410389590889856 &  369612136 &  208.52 &  10.89 &   -1.42 &   8.4 &  0.1 &  0.1 &  208.2 &  -27.8 &   17.4 &  255.6 \\
916214002815131520 &  496906047 &  177.57 &  34.46 &   -0.69 &   9.9 & -0.1 &  1.2 & -129.5 &  -43.9 &   26.5 &  443.9 \\
3892974510467742080 &  237709247 &  274.63 &  63.71 &   -0.30 &   8.2 &  0.5 &  1.1 &  219.4 &  -22.2 &  -43.1 &  295.2 \\
3844841563599413760 &  296304241 &  230.31 &  35.74 &   -1.30 &   9.2 &  1.2 &  1.2 &  175.6 &  -24.9 &  -38.8 &  442.9 \\
828272077823821184 &  442003208 &  162.29 &  49.36 &   -0.77 &  10.3 & -0.7 &  2.5 & -159.4 &  -34.6 &   19.7 &  460.0 \\
2117423618572646528 &  365016133 &   73.97 &  19.67 &   -1.62 &   7.6 & -2.1 &  0.8 & -233.1 &    6.5 &    5.7 &  446.8 \\
2748328170890349184 &  181312168 &  107.44 & -55.64 &   -1.66 &   8.5 & -0.8 & -1.3 &  188.9 &  -65.1 &   51.6 &  393.1 \\
804217031156736000 &  547602063 &  180.67 &  55.70 &   -1.95 &  10.8 &  0.0 &  3.9 &  -70.1 &  -36.2 &   37.0 &  389.6 \\
842617680909921408 &  397809151 &  152.03 &  57.83 &   -1.50 &   8.8 & -0.3 &  1.1 & -204.0 &  -30.2 &   11.7 &  333.5 \\
3098112983239720064 &  308415134 &  215.18 &  23.01 &   -1.23 &   9.0 &  0.6 &  0.5 & -173.3 &  -48.6 &   -6.0 &  336.2 \\
685285091757994624 &  187405157 &  205.90 &  39.41 &   -1.88 &  10.2 &  1.0 &  1.8 & -118.1 &  -44.1 &    4.8 &  335.2 \\
974643017084542336 &  182201020 &  171.87 &  25.01 &   -1.12 &  11.7 & -0.5 &  1.7 &   -6.1 &  -28.9 &  -11.8 &  342.4 \\
721601548505308928 &  448215132 &  214.04 &  59.20 &   -1.65 &   9.2 &  0.7 &  2.1 &  177.5 &  -21.0 &  -51.5 &  317.5 \\
1571249502467492992 &  152801087 &  125.40 &  61.99 &   -1.44 &   9.5 & -1.8 &  4.2 & -129.4 &   -6.2 &   71.3 &  292.4 \\
1492666547437694464 &  569710210 &   79.10 &  65.84 &   -1.14 &   8.0 & -1.1 &  2.6 & -218.2 &  -14.1 &   48.8 &  358.5 \\
1293408205528224512 &  457905178 &   58.69 &  63.05 &   -1.08 &   7.3 & -1.4 &  3.3 & -198.7 &   -1.1 &   77.2 &  288.4 \\
740747516278184576 &  339902204 &  199.71 &  54.45 &   -0.94 &   9.5 &  0.5 &  2.0 &  148.0 &  -33.6 &  -24.9 &  388.2 \\
39345031464898304 &  254103211 &  176.36 & -27.63 &   -0.83 &  10.7 & -0.2 & -1.3 &  153.5 &  -38.0 &   38.9 &  381.9 \\
884525394804288384 &  524204063 &  188.64 &  15.61 &   -1.86 &  11.7 &  0.5 &  1.0 &  -11.3 &  -29.3 &   13.8 &  336.5 \\
140259720489824768 &  496804234 &  149.43 & -21.54 &   -1.40 &  10.6 & -1.4 & -1.1 & -138.2 &  -12.7 &    1.5 &  329.3 \\
3916715165534162048 &  339504190 &  253.13 &  68.65 &   -0.61 &   8.3 &  0.5 &  1.3 & -210.7 &  -65.2 &   13.3 &  440.9 \\
836462443019621120 &  397803185 &  155.03 &  57.63 &   -1.19 &  10.1 & -0.9 &  3.4 & -134.0 &  -24.7 &   42.4 &  369.9 \\
1376562551949814912 &  153310184 &   60.97 &  50.43 &   -0.97 &   7.9 & -0.6 &  0.9 &  237.6 &  -65.7 &  -39.7 &  366.5 \\
578849205092501632 &  207312070 &  227.14 &  32.32 &   -1.36 &  11.2 &  3.2 &  2.8 &   16.2 &  -27.6 &   -6.7 &  360.9 \\
828352101654416640 &  442015248 &  161.53 &  49.31 &   -1.39 &  11.6 & -1.1 &  4.2 &  -24.2 &  -37.2 &    1.5 &  460.0 \\
3398415211089085440 &  258013114 &  189.27 &  -4.67 &   -0.99 &  11.8 &  0.6 & -0.3 &  -78.7 &  -36.2 &   12.6 &  381.8 \\
3072883525005327488 &  196812249 &  227.04 &  20.11 &   -1.09 &  11.5 &  3.6 &  1.8 &  -59.7 &  -46.0 &   16.2 &  316.7 \\
1309070576867596032 &  229601019 &   52.09 &  33.68 &   -1.34 &   6.3 & -2.4 &  2.1 &  185.0 & -114.0 &  -93.8 &  271.5 \\
1015712868757539712 &  327109055 &  169.84 &  41.52 &   -0.80 &  12.4 & -0.8 &  3.8 &  -11.6 &  -24.7 &    7.0 &  316.0 \\
3969220556612007168 &  331210090 &  232.65 &  60.73 &   -0.80 &   9.3 &  1.4 &  3.2 & -153.8 &  -75.2 &   63.5 &  479.8 \\
375106436290960384 &    7013119 &  123.48 & -21.25 &   -0.66 &  10.4 & -3.3 & -1.5 & -139.9 &    4.6 &   -5.5 &  416.2 \\
1544386131299665024 &  315612213 &  129.50 &  67.46 &   -0.61 &   9.0 & -1.0 &  3.1 & -167.6 &  -15.7 &   33.0 &  303.7 \\
4009546245072890368 &  299710019 &  211.31 &  82.49 &   -1.25 &   8.3 &  0.1 &  1.2 &  209.8 &  -29.5 &  -15.5 &  261.8 \\
3090248799337115648 &   21206245 &  221.57 &  21.10 &   -0.66 &  11.4 &  2.8 &  1.7 &  -40.5 &  -44.2 &    2.2 &  389.0 \\
\hline
\end{longtable}
\end{center}

\begin{center}
	\begin{longtable}{cccccccccccc}
\caption{GL-4, 75 stars}\\
\hline
\hline
source-id &      obsid &       l &      b &  [Fe/H] &     x &    y &     z &      U &      V &      W &     $L_z$     \\
&            & (deg)   & (deg) 　& (dex)   & (kpc) & (kpc)& (kpc) & (km s$^{-1}$) & (km s$^{-1}$) & (km s$^{-1}$) & (km s$^{-1}$ kpc) \\
\hline
 3862691074020171392 &  215114081 &  238.39 &  51.44 &   -1.14 &  8.5 &  0.5 &  0.7 &  -64.8 & -100.7 &  21.2 &   824.5 \\
3983060556147457920 &  341205081 &  225.63 &  62.22 &   -1.32 &  8.6 &  0.4 &  1.2 &   44.0 & -100.5 &   9.4 &   884.6 \\
3998046590396332544 &  219209044 &  204.45 &  70.14 &   -1.55 &  8.6 &  0.2 &  1.4 &  -38.3 & -105.2 &  10.7 &   901.4 \\
629560498136247680 &  491914225 &  211.52 &  52.11 &   -1.70 &  8.5 &  0.2 &  0.5 &   39.0 & -122.8 & -11.9 &  1057.0 \\
1030072353177238656 &  427411139 &  162.95 &  39.15 &   -1.28 &  8.7 & -0.1 &  0.4 &    3.1 & -113.3 &  16.5 &   980.4 \\
1281563304200564480 &  210407087 &   42.03 &  64.71 &   -1.39 &  7.8 & -0.4 &  1.3 &  -61.5 & -114.5 & -11.6 &   913.0 \\
1361967737682703744 &  346304069 &   72.93 &  32.18 &   -1.13 &  7.9 & -0.9 &  0.6 &  117.2 & -114.5 &  15.7 &   800.0 \\
1460408456826617216 &  301808195 &   28.02 &  87.53 &   -1.36 &  8.1 & -0.0 &  1.4 &  -53.8 &  -97.0 &  -6.7 &   791.9 \\
1589273109387131648 &  417415146 &   81.31 &  54.73 &   -1.27 &  8.1 & -0.5 &  0.7 &   10.1 & -109.3 &  24.9 &   883.2 \\
4501173868500580224 &  448015177 &   40.28 &  21.46 &   -0.73 &  5.4 & -2.4 &  1.5 &   88.4 & -187.3 & -38.8 &   807.0 \\
237446274827231104 &  372308145 &  153.13 & -10.71 &   -1.29 &  8.7 & -0.3 & -0.1 &  -66.8 & -108.0 &  11.4 &   960.3 \\
54587801677441280 &  406512088 &  167.30 & -32.80 &   -1.44 &  8.6 & -0.1 & -0.2 &  -95.5 & -111.5 & -13.0 &   965.7 \\
2051842324124265984 &  369807060 &   70.83 &   9.22 &   -1.52 &  7.3 & -2.7 &  0.5 & -131.2 &  -65.5 & -20.8 &   831.5 \\
2092283667462120960 &  574205190 &   64.20 &  14.55 &   -1.33 &  6.5 & -3.6 &  1.1 &   56.3 & -168.0 &  -7.9 &   883.8 \\
2106705205566435200 &  457213047 &   74.46 &  17.51 &   -1.19 &  7.6 & -2.2 &  0.8 & -163.7 &  -62.7 &  20.9 &   837.7 \\
2550420510295875712 &  250203025 &  118.62 & -60.13 &   -1.40 &  8.4 & -0.3 & -0.7 & -112.1 & -105.9 &  12.4 &   927.2 \\
2567974041634213760 &  473508208 &  149.06 & -52.56 &   -1.54 &  8.9 & -0.4 & -1.0 &   78.3 &  -99.5 &  24.9 &   850.8 \\
2758035174935654784 &  361515200 &   94.37 & -51.00 &   -1.37 &  8.3 & -0.7 & -0.8 &  -72.2 & -106.0 & -28.1 &   922.7 \\
2828778027944351488 &  363008015 &   85.46 & -37.52 &   -1.39 &  8.0 & -2.6 & -2.0 &   36.9 & -111.3 &   9.1 &   792.8 \\
2864023182410285824 &  284614066 &  115.54 & -28.48 &   -1.65 &  8.6 & -0.9 & -0.5 &  -12.2 &  -94.6 &  -8.3 &   825.5 \\
2864505043382651136 &  164509075 &  103.03 & -34.32 &   -1.37 &  8.7 & -2.0 & -1.4 &  -16.3 &  -87.5 &   4.1 &   791.1 \\
3381170883031960192 &  401003217 &  191.03 &   9.95 &   -1.26 &  8.7 &  0.1 &  0.1 &  -44.9 & -108.1 &  18.0 &   937.9 \\
397548396328513024 &  182713017 &  128.26 & -18.33 &   -1.37 &  8.3 & -0.1 & -0.0 &  105.3 & -115.3 &  -4.0 &   943.2 \\
3674143933870596352 &  339608084 &  349.10 &  62.97 &   -1.82 &  7.9 &  0.0 &  0.5 &  127.2 & -112.4 & -32.0 &   899.2 \\
3701146668138963840 &  435612027 &  281.27 &  63.51 &   -1.05 &  8.1 &  0.3 &  0.7 &  114.1 & -102.2 &   7.4 &   866.7 \\
3701052316297428992 &   94114192 &  283.66 &  63.96 &   -1.15 &  8.1 &  0.6 &  1.3 &  -33.4 & -117.2 &  24.5 &   924.9 \\
3819896977659791104 &   83407024 &  244.34 &  33.49 &   -1.64 &  8.8 &  1.2 &  0.9 &   37.5 & -109.4 &  -2.7 &  1003.4 \\
3839477875296667904 &  408915165 &  236.23 &  34.94 &   -1.37 &  8.5 &  0.5 &  0.5 &   65.2 &  -93.9 & -16.7 &   835.4 \\
3867919664127132928 &  531306171 &  240.03 &  58.23 &   -1.19 &  8.4 &  0.3 &  0.6 &   30.0 & -123.9 &  11.4 &  1044.9 \\
3945839785445446784 &  202016139 &  274.01 &  78.04 &   -1.46 &  8.2 &  0.1 &  0.4 &   -7.6 & -115.9 &  14.9 &   949.3 \\
3957868545732319360 &  334509203 &  293.47 &  87.45 &   -0.92 &  8.2 &  0.0 &  1.2 &  -86.5 & -111.0 &  -2.1 &   903.8 \\
3983648794868003456 &  215807088 &  227.66 &  63.96 &   -1.49 &  8.3 &  0.2 &  0.5 &   80.9 &  -98.6 &  27.5 &   836.3 \\
4011197333580540416 &  153508020 &  152.79 &  86.86 &   -1.31 &  8.3 & -0.0 &  1.6 &   78.2 & -105.3 & -31.1 &   868.7 \\
4013781873100372992 &  508012079 &  187.43 &  81.20 &   -1.53 &  8.3 &  0.0 &  0.9 &   28.0 & -125.5 & -24.2 &  1046.9 \\
4028191728896688128 &  399112211 &  180.33 &  76.66 &   -1.36 &  8.5 &  0.0 &  1.1 &   34.9 & -114.3 &   2.0 &   967.7 \\
4029890852318475392 &  556204185 &  171.58 &  76.86 &   -1.43 &  8.3 & -0.0 &  0.5 &  -51.1 & -109.6 &  20.1 &   912.2 \\
95253441093777408 &  378502232 &  140.92 & -41.85 &   -1.19 &  8.7 & -0.4 & -0.5 &  -48.2 & -110.9 & -26.9 &   980.6 \\
15665777248819968 &  368607203 &  169.27 & -38.46 &   -2.25 &  9.3 & -0.2 & -0.8 &   16.4 & -107.1 &  20.7 &   987.3 \\
676481130355532032 &  215006143 &  201.11 &  28.03 &   -1.43 &  8.6 &  0.2 &  0.3 &   12.6 & -106.9 & -24.4 &   921.2 \\
723727415582400768 &  523308121 &  210.91 &  61.52 &   -1.18 &  8.6 &  0.2 &  0.8 &   57.1 &  -87.6 & -31.0 &   763.9 \\
788860014820041216 &  563004050 &  155.21 &  63.94 &   -0.93 &  8.6 & -0.2 &  0.9 &  -42.8 & -101.9 &  14.9 &   882.8 \\
893434119048333184 &  226714179 &  185.56 &  21.97 &    0.23 &  8.4 &  0.0 &  0.1 &  -21.1 & -110.2 &  -7.4 &   929.3 \\
933175245278987520 &  339206053 &  170.83 &  30.58 &   -1.05 &  8.7 & -0.1 &  0.3 &  -68.0 &  -96.1 &  -3.8 &   838.4 \\
1284876441972634752 &  207004216 &   46.36 &  69.89 &   -1.54 &  7.8 & -0.4 &  1.7 &  102.2 & -116.2 & -30.0 &   861.2 \\
1358501252397563136 &  150401057 &   69.70 &  37.55 &   -1.43 &  7.6 & -1.5 &  1.3 &   48.1 & -144.3 &  -9.7 &  1026.2 \\
1377857192531501952 &  458402167 &   64.13 &  52.35 &   -1.26 &  7.9 & -0.6 &  0.8 & -101.6 & -105.6 &   7.8 &   894.3 \\
1382300215940599296 &  576315040 &   68.86 &  44.09 &   -1.31 &  7.9 & -0.9 &  0.9 &  -64.5 & -101.1 &   8.4 &   851.7 \\
1387001173609203584 &  433902063 &   73.89 &  45.61 &   -1.38 &  7.6 & -2.0 &  2.1 &  -82.8 &  -82.8 &  17.9 &   794.2 \\
1403723684611249536 &  437204196 &   79.83 &  47.74 &   -1.42 &  8.1 & -0.6 &  0.7 &  -67.2 & -123.3 & -21.3 &  1038.9 \\
1457430567021692416 &  546609055 &   57.84 &  76.38 &   -1.29 &  8.1 & -0.2 &  0.9 &   55.8 & -106.2 & -29.6 &   848.9 \\
1461674063428369920 &   44106194 &   49.20 &  82.96 &   -1.40 &  8.2 & -0.0 &  0.2 &   -8.8 & -110.9 & -10.2 &   908.1 \\
1477755554856192896 &  567811131 &   51.59 &  69.83 &   -1.37 &  8.1 & -0.1 &  0.4 &   82.5 & -132.7 & -19.2 &  1067.6 \\
1491879194032720512 &  455609065 &   75.59 &  66.74 &   -1.30 &  8.1 & -0.4 &  1.1 &   -5.2 & -132.3 & -21.8 &  1072.7 \\
1545676545638836096 &  448503076 &  139.33 &  67.71 &   -1.92 &  8.5 & -0.3 &  1.1 &   59.1 & -101.6 &  17.5 &   849.6 \\
1578299639743871488 &  438711142 &  125.82 &  57.86 &   -1.35 &  8.4 & -0.3 &  0.6 & -112.4 & -109.6 &  -3.1 &   959.5 \\
1589188687509570048 &  417416161 &   82.65 &  55.68 &   -1.48 &  8.2 & -0.2 &  0.4 & -119.3 & -105.1 &  15.8 &   885.4 \\
1795085983706056960 &  162916207 &   78.54 & -23.52 &   -1.12 &  8.0 & -1.0 & -0.4 &  -59.2 & -110.0 & -26.1 &   938.8 \\
1215531068242641408 &  239005015 &   34.69 &  55.58 &   -0.83 &  7.6 & -0.4 &  1.1 &   45.8 & -134.8 & -29.5 &  1001.1 \\
747669354292137472 &  446515198 &  198.26 &  58.59 &   -2.24 &  9.0 &  0.3 &  1.4 &   29.9 &  -92.9 &  20.6 &   842.4 \\
4010300132092253440 &  450616195 &  196.27 &  82.73 &   -0.92 &  8.4 &  0.0 &  1.3 &    4.3 & -112.0 &   2.3 &   935.9 \\
105468733044869888 &  173205016 &  143.56 & -33.87 &   -1.29 &  8.6 & -0.3 & -0.3 &   -6.2 & -115.6 & -15.7 &   994.1 \\
647812945207911040 &   19104199 &  200.23 &  47.07 &   -1.23 &  8.6 &  0.1 &  0.5 &   28.4 &  -97.1 & -26.2 &   838.5 \\
2131873984499709312 &  249214199 &   78.90 &  19.75 &   -1.30 &  7.3 & -4.7 &  1.8 &  -50.7 &  -94.4 &  10.5 &   926.1 \\
298802390907012992 &  267312041 &  137.82 & -33.54 &   -1.51 &  8.7 & -0.5 & -0.4 &    2.4 & -114.8 &  14.5 &  1000.6 \\
3727921215088631296 &  228311012 &  346.20 &  69.52 &   -1.54 &  7.7 &  0.1 &  1.4 &   75.4 & -112.8 &   7.2 &   876.4 \\
3985817787712703872 &  106004185 &  223.11 &  58.70 &   -1.40 &  8.6 &  0.4 &  1.0 &   64.3 &  -91.9 & -10.2 &   817.9 \\
1264626667604880896 &  435811144 &   36.27 &  59.49 &   -1.43 &  7.6 & -0.4 &  1.2 & -124.1 & -112.3 &  19.0 &   908.6 \\
2864663682294138496 &  359806154 &  101.46 & -33.63 &   -1.48 &  8.6 & -2.1 & -1.4 &  -84.6 &  -72.4 &  13.7 &   805.3 \\
736762125930364032 &  505504085 &  196.24 &  63.74 &   -1.25 &  8.8 &  0.2 &  1.4 &   93.1 &  -95.2 &  -8.8 &   859.4 \\
888198313397015552 &  134516227 &  186.50 &  15.43 &    0.00 &  9.4 &  0.1 &  0.3 &    2.1 &  -98.9 &  10.6 &   925.5 \\
2051627129085490944 &  369807223 &   70.38 &   8.78 &   -1.62 &  6.1 & -6.0 &  1.0 & -120.6 &  -28.8 &  22.1 &   898.4 \\
1522063635777101568 &  419909082 &  131.20 &  76.34 &   -1.22 &  8.4 & -0.2 &  1.2 &  -68.3 & -113.1 &   6.8 &   962.6 \\
4010006631207125504 &  299815052 &  201.38 &  83.49 &   -1.32 &  8.3 &  0.0 &  0.9 &   59.3 & -104.9 & -13.1 &   871.5 \\
832771073246198016 &  385716075 &  158.71 &  59.47 &   -1.08 &  8.3 & -0.1 &  0.3 &   96.9 & -101.8 &  25.6 &   843.5 \\
3892473201883627648 &  457705242 &  271.78 &  61.90 &   -1.37 &  8.2 &  0.4 &  0.8 &  -17.7 & -105.2 &  25.9 &   854.2 \\
\hline
\end{longtable}
\end{center}



\begin{thebibliography}{}


\bibitem[Abadi et al.(2003)]{Abadi03}
Abadi, M. G., Navarro, J. F., \& Steinmetz, M. 2003, ApJ, 597, 21

\bibitem[Allende Prieto et al.(2006)]{Allende Prieto06}
Allende Prieto, C., Beers, T. C., Wilhelm, R., et al. 2006, ApJ, 636, 804

\bibitem[Antoja et al.(2008)]{Antoja08}
Antoja, T., Figueras, F., Fernandez, D., et al. 2008, A\&A, 490, 135

\bibitem[Antoja et al.(2012)]{Antoja12}
Antoja, T., Helmi, A., Bienayme, O., et al. 2012, MNRAS, 426, L1

\bibitem[Bailer-Jones et al.(2018)]{Bailer18}
Bailer-Jones, C. A. L., Rybizki, J., Fouesneau, M., et al., 2018, AJ, 156, 58

\bibitem[Belokurov et al.(2018)]{Belokurov18}
Belokurov, V., Erkal, D., Evans, N. W., et al. 2018, MNRAS, 478, 611 

\bibitem[Belokurov et al.(2007)]{Belokurov07}
Belokurov, V., Evans, N. W., Irwin, M. J., et al. 2007, ApJ, 658, 337 

\bibitem[Belokurov et al.(2006)]{Belokurov06}
Belokurov, V., Zucker, D. B., Evans, N. W., et al. 2006, ApJ Lett, 642, L137

\bibitem[Bensby et al.(2003)]{Bensby03}
Bensby, T., Feltzing, S., \& Lundström, I. 2003, A\&A, 410, 527

\bibitem[Bland-Hawthorn \& Gerhard (2016)]{Bland16}
Bland-Hawthorn, J., \& Gerhard, O. 2016, ARA\&A, 54, 529

\bibitem[Bonaca et al.(2017)]{Bonaca17}
Bonaca, A., Conroy, C., Wetzel, A., et al. 2017, ApJ, 845, 101

\bibitem[Bond et al.(2010)]{Bond10}
Bond, N. A., Ivezi\'c, \v{Z}, Sesar, B., et al. 2010, ApJ 716, 1

\bibitem[Bullock \& Johnston (2005)]{Bullock05}
Bullock, J. S., \& Johnston, K. V. 2005, ApJ, 635, 931

\bibitem[Carollo et al.(2007)]{Carollo07}
Carollo, D., Beers, T. C., Lee, Y. S., et al. 2007, Nature, 450, 1020

\bibitem[Carollo et al.(2010)]{Carollo10}
Carollo, D., Beers, T. C., Chiba, M., et al. 2010, ApJ, 712, 692

\bibitem[Chiba \& Beers (2000)]{Chiba00}
Chiba, M., \& Beers, T. C. 2000, AJ, 119, 2843

\bibitem[Cooper et al.(2015)]{Cooper15}
Cooper, A. P., Parry, O. H., Lowing, B., et al. 2015, MNRAS, 454, 3185

\bibitem[Cui et al.(2012)]{Cui12}
Cui, X. Q., Zhao, Y. H., Chu, Y. Q., et al. 2012, RAA, 12, 1197

\bibitem[Deason et al.(2017)]{Deason17}
Deason, A. J., Belokurov, V., Koposov S. E., et al. 2017, MNRAS, 470, 1259

\bibitem[Dehnen \& Binney (1998)]{Dehnen98}
Dehnen, W., \& Binney, J. J. 1998, MNRAS, 298, 387

\bibitem[Deng et al.(2012)]{Deng12}
Deng, L. C., Newberg, H. J., Liu C., et al. 2012, RAA, 12, 735

\bibitem[Diemand et al.(2008)]{Diemand08}
Diemand, J., Kuhlen, M., Madau, P., et al. 2008, Nature, 454, 735

\bibitem[Duffau et al.(2014)]{Duffau14}
Duffau, S., Vivas, A. K., Zinn, R., et al. 2014, A\&A, 566, A118

\bibitem[Duffau et al.(2006)]{Duffau06}
Duffau, S., Zinn, R., Vivas, A. K., et al. 2006, ApJ Lett, 636, L97


\bibitem[Eggen et al.(1962)]{Eggen62}
Eggen, O. J., Lynden-Bell, D., \& Sandage, A. R. 1962, ApJ, 136, 748

\bibitem[Famaey et al.(2005)]{Famaey05}
Famaey, B., Jorissen, A., Luri, X., et al. 2005, A\&A, 430, 165

\bibitem[Fern\'{a}ndez-Trincado et al.(2015)]{Fernandez15}
Fern\'{a}ndez-Trincado, J. G., Robin, A. C., Vieira, K., et al. 2015, A\&A, 583, A76

\bibitem[Font et al.(2011)]{Font11}
Font, A. S., Mccarthy, I. G., Crain, R. A., et al. 2011, MNRAS, 416, 2802

\bibitem[Freeman \& Bland-Hawthorn (2002)]{Freeman02}
Freeman, K., \& Bland-Hawthorn, J. 2002, ARA\&A, 40, 487

\bibitem[Gaia Collaboration (2018)]{Gaia18}
Gaia Collaboration, (Brown, A. G. A., et al.) 2018, A\&A, 616, A1

\bibitem[Grillmair \& Carlin (2016)]{Grillmair16}
Grillmair, C. J., \& Carlin, J. L. 2016, ASSL, 420, 87

\bibitem[Grillmair(2006)]{Grillmair06}
Grillmair, C. J.\ 2006, ApJ Lett, 645, L37 

\bibitem[Helmi et al.(2017)]{Helmi17}
Helmi, A., Veljanoski J., Breddels M. A., et al. 2017, A\&A , 598, A58

\bibitem[Helmi \& de Zeeuw(2000)]{Helmi00}
Helmi, A., \& de Zeeuw, P. T. 2000, MNRAS, 319, 657

\bibitem[Helmi \& White(1999a)]{Helmi99a}
Helmi, A., \& White, S. D. M. 1999a, MNRAS, 307, 495

\bibitem[Helmi et al.(1999b)]{Helmi99b}
Helmi, A., White S. D. M., de Zeeuw P. T., et al. 1999b, Nature, 402, 53

\bibitem[Huang et al.(2018)]{Huang18}
Huang, Y., Liu, X. W., Chen, B. Q., et al. 2018, AJ, 156, 90

\bibitem[Ibata et al.(1994)]{Ibata94}
Ibata, R. A., Gilmore, G., Irwin, M. J. 1994, Nature, 370, 194


\bibitem[Johnson \& Soderblom(1987)]{Johnson87}
Johnson, D. R. H., \& Soderblom, D. R. 1987, AJ, 93, 864

\bibitem[Juri\'c et al.(2008)]{Juric08}
Juri\'c M., Ivezi\'c, \v{Z}., Brooks, A., et al. 2008, ApJ, 673, 864

\bibitem[Kepley et al.(2007)]{Kepley07}
Kepley, A. A., Morrison, H. L., Helmi, A., et al. 2007, AJ, 134, 1579

\bibitem[Klement et al.(2008)]{Klement08}
Klement, R., Fuchs, B., \& Rix, H. W. 2008, ApJ, 685, 261

\bibitem[Klement (2010)]{Klement10}
Klement, R. J. 2010, A\&A Rev., 18, 567

\bibitem[Klement et al.(2009)]{Klement09}
Klement, R., Rix, H. W, Flynn, C., et al. 2009, ApJ, 698, 865

\bibitem[Klypin et al.(2011)]{Klypin11}
Klypin, A. A., Trujillo-Gomez, S., \& Primack, J. 2011, ApJ, 740, 102

\bibitem[Kordopatis et al.(2013)]{Kordopatis13}
Kordopatis, G., Gilmore, G., Wyse, R. F. G., et al. 2013, MNRAS, 436, 3231

\bibitem[Li et al.(2017)]{Li17}	
Li, G. W., Yanny, B., Zhang, H. T., et al. 2017, RAA, 17, 62

\bibitem[Liang et al.(2017)]{Liang17}
Liang, X. L., Zhao, J. K., Oswalt, T. D., et al., 2017, ApJ, 844, 152

\bibitem[Lindegren et al.(2018)]{Lindegren18}
Lindegren, L., Hern$\acute{\mathrm{a}}$ndez, J., Bombrun, A., et al. 2018, AJ, 616, A2

\bibitem[Lindegren (2016)]{Lindegren16}
Lindegren, L., Lammers, U., Bastian, U., et al. 2016, A\&A, 595, A4

\bibitem[Luri et al.(2018)]{Luri18}
Luri, X., Brown, A. G. A., Sarro, L. M., et al. 2018, A\&A, 616, A9

\bibitem[Luo et al.(2015)]{Luo15}
Luo, A. L., Zhao, Y. H., Zhao, G., et al. 2015, RAA, 15, 1095

\bibitem[Majewskia et al.(2003)]{Majewski03}
Majewski, S. R., Skrutskie, M. F., Weinberg, M. D., et al. 2003, ApJ, 599, 1082

\bibitem[McMillan (2017)]{McMillan17}
McMillan, P. J., 2017, MNRAS, 465, 76

\bibitem[Morrison et al.(2009)]{Morrison09}
Morrison, H. L., Helmi, A., Sun, J.-Y., et al. 2009, ApJ, 694, 130

\bibitem[Newberg et al.(2010)]{Newberg10}
Newberg, H. J., Willett, B. A., Yanny, B., \& Xu, Y.  2010, ApJ, 711, 32 

\bibitem[Newberg et al.(2007)]{Newberg07}
Newberg, H. J., Yanny, B., Cole, N., et al. 2007, ApJ, 668, 221

\bibitem[Pedregosa et al.(2012)]{Pedregosa12}
Pedregosa, F., Varoquaux, G., Gramfort, A., et al. 2012, arXiv:1201.0490

\bibitem[Re Fiorentin et al.(2015)]{Re Fiorentin15}
Re Fiorentin, P., Lattanzi, M. G., Spagna, A., et al. 2015, AJ, 150, 128

\bibitem[Sales et al.(2007)]{Sales07}
Sales, L. V., Navarro, J. F., Abadi, M. G., et al. 2007, MNRAS, 379, 1464

\bibitem[Searle \& Zinn(1978)]{Searle78}
Searle, L., \& Zinn, R. 1978, ApJ, 225, 357

\bibitem[Skrutskie et al.(1997)]{2MASS}
Skrutskie, M. F., Schneider, S. E., Stiening, R., et al. 1997, ASSL, 210, 25 

\bibitem[Smith et al.(2009)]{Smith09}
Smith, M. C., Evans, N. W., Belokurov, V., et al. 2009, MNRAS, 399, 1223

\bibitem[Springel et al.(2008)]{Springel08}
Springel, V., Wang, J., Vogelsberger, M., et al. 2008, MNRAS, 391, 1685 

\bibitem[Steinmetz et al.(2006)]{Steinmetz06}
Steinmetz, M., Zwitter, T., Siebert, A., et al. 2006, AJ, 132, 1645

\bibitem[Tissera et al.(2014)]{Tissera14}	
Tissera, P. B., Beers, T. C., Carollo, D., et al. 2014, MNRAS, 439, 3128

\bibitem[Tian et al.(2015)]{Tian15}
Tian, H. J., Liu, C., Carlin, J. L., et al. 2015, ApJ, 809, 145

\bibitem[van der Walt et al.(2014)]{Walt14}
van der Walt, S., Sch$\ddot{\mathrm{o}}$nberger, J. L., Nunez-Iglesias, J., et al. 2014, PeerJ, 2, 453

\bibitem[van Leeuwen(2007)]{Leeuwen07}
van Leeuwen, F. 2007, ASSL, 350

\bibitem[Vincents \& Soille(1991)]{Vincents91}
Vincent, L., \& Soille, P. 1991, KvanE, 13, 583

\bibitem[Weiss \& Kulikowski (1991)]{Weiss91}
Weiss, S., \& Kulikowski, C. 1991, Computer Systems that Learn: Classification and Prediction Methods from Statistics, Neural Nets, Machine Learning, and Expert Systems

\bibitem[Wright et al.(2010)]{WISE}
Wright, E. L., Eisenhardt, P. R. M., Mainzer, A. K., et al. 2010, AJ, 140, 1868

\bibitem[Wu et al.(2011)]{Wu11}
Wu, Y., Luo, A. L., Li, H. N., et al. 2011, RAA, 11, 924

\bibitem[York et al.(2000)]{York00}
York, D. G., Adelman, J., Anderson, J. E., Jr., et al. 2000, AJ, 120, 1579 

\bibitem[Zhao et al.(2012)]{Zhao12}
Zhao, G., Zhao, Y.-H., Chu, Y.-Q., et al. 2012, RAA, 12, 723

\bibitem[Zhao et al.(2009)]{Zhao09}
Zhao, J. K., Zhao, G., \& Chen, Y. Q. 2009, ApJ Lett, 692, L113

\bibitem[Zhao et al.(2014)]{Zhao14}
Zhao, J. K., Zhao, G., Chen, Y. Q., et al. 2014, ApJ, 787, 31

\bibitem[Zhao et al.(2015)]{Zhao15}
Zhao, J. K., Zhao, G., Chen, Y. Q., et al. 2015, RAA, 15, 1378

\bibitem[Zolotov et al.(2010)]{Zolotov10}
Zolotov, A., Willman, B., Brooks, A., et al. 2010, ApJ, 721, 738

\bibitem[Zuo et al.(2017)]{Zuo17}
Zuo, W.B., Du, C.H., Jing, Y.J., et al. 2017, ApJ, 841, 59

\end{thebibliography}
\end{document}